\documentclass[aps,prb,superscriptaddress]{revtex4}
\usepackage{graphicx,bm,amssymb,color} 
\usepackage{amsfonts} 
\begin{document} 
\title{Spin-orbit interaction and anomalous spin relaxation \\
in carbon nanotube quantum dots} 
\author{Denis V. Bulaev}
\affiliation{Department of Physics, University of Basel, Klingelbergstrasse 82, CH-4056 Basel, Switzerland}
\author{Bj\"orn Trauzettel}
\affiliation{Department of Physics, University of Basel, Klingelbergstrasse 82, CH-4056 Basel, Switzerland}
\affiliation{Institute of Theoretical Physics and Astrophysics, University of W\"urzburg, D-97074 W\"urzburg, Germany}
\author{Daniel Loss}
\affiliation{Department of Physics, University of Basel, Klingelbergstrasse 82, CH-4056 Basel, Switzerland}

\begin{abstract}
We study spin relaxation and decoherence in nanotube quantum dots caused by electron-lattice and
spin-orbit interaction and predict striking effects induced by magnetic fields
$B$. For particular values
of $B$, destructive interference occurs resulting in ultralong spin
relaxation  times $T_1$ exceeding tens of seconds. For small phonon frequencies $\omega$, we find a
$1/\sqrt{\omega}$ spin-phonon noise spectrum --
a dissipation channel for spins in quantum dots -- which can reduce
$T_1$ by many orders of magnitude. We show that
nanotubes exhibit zero-field level splitting caused by spin-orbit
interaction. This enables an all-electrical and phase-coherent control of
spin.
\end{abstract}

\maketitle
\section{Introduction}
Although semiconductor spintronics is a field with already a substantial
history as well as with commercial applications,\cite{Awsch07} spintronics with
carbon-based materials is a young research area with excellent perspectives.
Only very recently, a pronounced gate-controlled magnetoresistance
response in carbon nanotubes connected to ferromagnetic leads has been
reported.\cite{Sahoo05} Furthermore, spin injection and detection in single-wall
carbon nanotubes has been demonstrated using a four-terminal 
geometry.\cite{Tombr06} The interest to implement spintronic devices
with carbon materials such as
carbon nanotubes \cite{Ando2005} or graphene \cite{Geim07}
is mainly driven by the desire to improve material
properties, for instance, for the spin relaxation behavior in these materials (as
compared to more standard semiconductors like GaAs). This is so because
carbon
is a comparably light atom, thus, spin-orbit interaction is typically
weak.\cite{Min06}
Additionally, it consists predominantly of $^{12}$C, which has zero nuclear
spin, thus, spin decoherence and relaxation caused by the hyperfine
interaction of the electron spin with the surrounding nuclear spins 
is weak. The advantageous material properties of carbon also trigger a large
interest to create spin qubits \cite{Loss98} in such materials.
%Recently, we have proposed
%a structure to form spin qubits in graphene \cite{TBLB2007}. This is a
%non-trivial task because one has to find a way to confine particles in
%graphene and (at the same time) get rid of the valley degeneracy that comes
%from its peculiar bandstructure \cite{Geim07}.  We face a similar problem
%with carbon nanotubes which can be thought of as graphene rolled up into a
%cylinder. The resulting periodic boundary conditions yield that in some
%cases
%single wall carbon nanotubes are metallic and in other cases they are
%semiconducting \cite{Ando2005}. However, the valley degeneracy is in
%general
%not lifted which makes it hard to do two-qubit gate operations in a double
%dot structure \cite{Burka99}.
%We show below that, in nanotubes, a magnetic field applied along the tube
%axis
%allows us to increase the spin relaxation
%time substantially.

Here, we provide quantitative calculations of spin relaxation and
spin decoherence times and show that they are dominated by a combination of
spin-orbit  and electron-phonon interaction.
It turns out that such spin-orbit induced effects get
 strongly  enhanced in small-radius nanotubes due to the curvature of the
lattice, and result in energy splittings that even exceed those occurring in
GaAs nanostructures.

The interplay of such enhanced spin-orbit interaction with the
one-dimensional nature of  nanotubes results in
a complex behavior with an extremely wide range of
relaxation rates which can be varied over many orders of magnitude by an
external magnetic field applied along the tube axis.
We show that interference effects can result in {\it
  ultralong} spin relaxation times exceeding tens of seconds. By contrast,
we uncover that for nanotube quantum dots, a spin-phonon
dissipation channel exists with a sub-Ohmic
spectral function ($\propto 1/\sqrt{\omega}$, see below) which results in {\it
 decreasing} spin relaxation times for decreasing spin level splitting $\omega$. Compared
to standard quantum dots (such as  GaAs or InAs semiconductors)
this is a most surprising behavior, since usually the spin decay
times increase  for decreasing
$\omega$.\cite{Khaet00,GKL,BLPRB05}

Most
remarkably, at zero magnetic field, the spin-orbit interaction
induces a zero-field splitting in the energy spectrum. We show that this
opens the door for an all-electrical control of spin in nanotube quantum
dots, again based on the strong spin-orbit interaction. This feature is most
interesting for spintronics applications where one aims at a spin
manipulation without making use of magnetic fields.
 Since quantum dots in
semiconducting carbon nanotubes have been realized by several
groups,\cite{Mason04,Sapma06,Graeb06,Tans,Bockrath,Kong,Minot,Pablo} we
believe that our
predictions are well within experimental reach.
%readily observable in the laboratory.

The paper is organized as follows. In~Section~\ref{sec:TeorModel}, we introduce a theoretical model for a nanotube quantum dot and solve the spectral problem of the Hamiltonian of such a system. In~Section~\ref{sec:SOI}, we study spin-orbit coupling in nanotubes, consider different contributions to the spin-orbit coupling, and investigate zero-field-level splitting induced by spin-orbit coupling.
In~Section~\ref{sec:ElPh}, electron-phonon coupling in nanotubes is considered. Analitical expressions for the coupling of an elecron to three deformational accoustic phonon modes are obtained. In~Section~\ref{sec:T1}, spin relaxation of an electron in a nanotube quantum dot is investigated and discussed.

\section{Theoretical model}
\label{sec:TeorModel}

\begin{figure}[thb]
\begin{center}
\includegraphics[width=10cm]{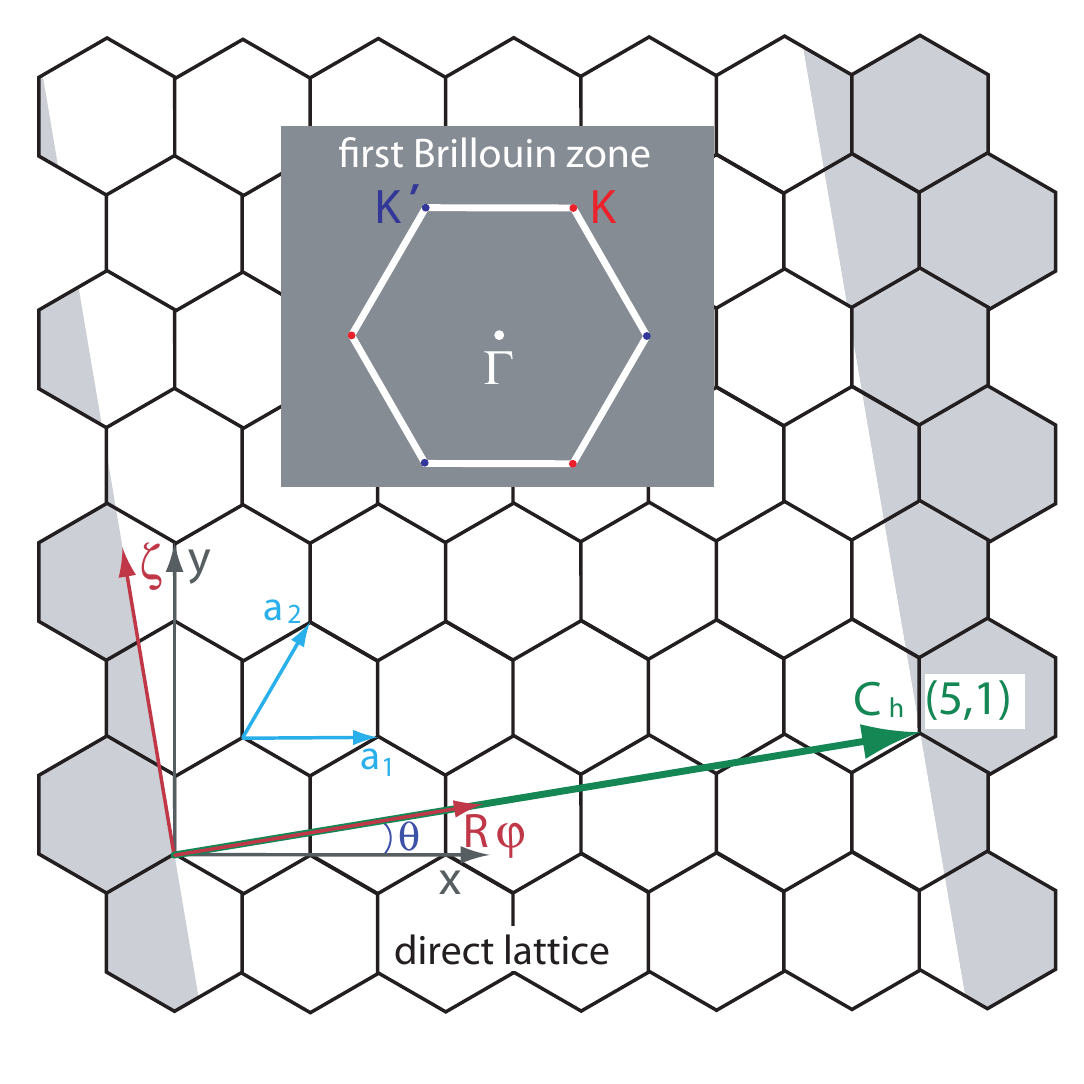}
\end{center}
\caption{ { Two-dimensional hexagonal lattice.} Here $\mathbf{a}_1$ and $\mathbf{a}_2$ are the primitive lattice vectors, $\mathbf{C}_h=n_1\mathbf{a}_1+n_2\mathbf{a}_2$ is the chiral vector (in this figure, we show the chiral vector with $n_1=5$ and $n_2=1$), $\theta$ is the chiral angle, $\zeta$ is along the NT axis, and $R\varphi$ is the azimuthal direction of a NT. In gray inset, the first Brillouin zone is depicted, where $\Gamma=(0,0)$ is the center of the zone; $\mathbf{K}$ and $\mathbf{K}'$ are non equivalent points in the Brillouin zone. }
\label{fig:SI_lattice}
\end{figure}

We consider a single wall nanotube (NT) defined by the chiral vector
$\mathbf{C}_h=n_1\mathbf{a}_1+n_2\mathbf{a}_2$, where $\mathbf{a}_1=a_0(1,0)$
and $\mathbf{a}_2=a_0(1/2,\sqrt{3}/2)$ are the primitive lattice vectors
($a_0=0.246\:$nm) and $n_1,n_2\in\mathbb{Z}$.\cite{Ando2005} The indices
$(n_1,n_2)$ determine the radius of a NT
$R=|\mathbf{C}_h|/2\pi=a_0\sqrt{n_1^2+n_2^2+n_1n_2}/2\pi$ and the chiral angle
(direction angle of $\mathbf{C}_h$) $\theta=\arctan[\sqrt{3}n_2/(2n_1+n_2)]$
(see Fig.~\ref{fig:SI_lattice}). 
 Neglecting curvature effects (which lead to an inessential shift of the valley
 minima in $\mathbf{k}$-space \cite{Ando2005}) and SOI, we
 describe the system at the $\mathbf{K}=(2\pi/a_0)(1/3,1/\sqrt{3})$ and the
 $\mathbf{K}'=(2\pi/a_0)(-1/3,1/\sqrt{3})$ point of the Brillouin zone (see Inset in Fig.~\ref{fig:SI_lattice}) by
 the Hamiltonian of graphene:\cite{DiVincenzo1984}
 \begin{eqnarray}%\nonumber
\tilde{H}_0&=&  \hbar v (\tau_3k_x\sigma_1+k_y\sigma_2)
= \hbar v \left( \begin{array}{cc}
0 & (\tau_3\kappa - i k)e^{-i\tau_3\theta}\\
(\tau_3\kappa + i k)e^{i\tau_3\theta} & 0\\
\end{array}\right),
\label{eq:H0tilde}
 \end{eqnarray}
 where $v$ is the Fermi velocity in a NT
 ($v=8.1\times10^{7}\:$cm$/$s)\cite{Lemay2001}, $\sigma_j$ are Pauli matrices
 operating on sublattice space, and  $\tau_3=1$ ($\tau_3=-1$) for the $\mathbf{K}$
 ($\mathbf{K}'$) point, $k$ is the electron wave-vector component along $\zeta$ and, $\kappa$ is along $\mathbf{C}_h$  (see Fig.~\ref{fig:SI_lattice}).
 It is convenient to perform a unitary transformation to remove  the dependence on the chirality angle $\theta$ from the Hamiltonian, i.e.:
 \begin{eqnarray}
\label{eq:UnTransform}
U&=& \left(  \begin{array}{cc}
e^{i\tau_3\theta} & 0 \\
0 & 1\\
\end{array} \right),\\
\label{eq:H0}
H_0&=& U\tilde{H_0}U^{-1}=\hbar v (\tau_3\kappa\sigma_1+k\sigma_2).
 \end{eqnarray}
 Eigenvalues and eigenfunctions (in the rotated reference frame $(\varphi,\zeta)$) of the Hamiltonian (\ref{eq:H0}) at zero magnetic field are given by 
 \begin{eqnarray}
\label{eq:Eigenvalue}
E_{\kappa,k}&=& \pm\hbar v\sqrt{\kappa^2+k^2},\\
\label{eq:Eigenfunction}
\Psi_{\kappa,k}^{(\prime)}(\varphi,\zeta)&=&  \frac{e^{i\mathbf{K^{(\prime)}}\cdot{\mathbf r}}}{\sqrt{4\pi}}e^{i(\kappa R\varphi+k\zeta)}
\left(  \begin{array}{c}z^{(\prime)}_{\kappa_m,k}\\
1\end{array}\right),\\
z_{\kappa,k}&=& \pm\frac{(\kappa-ik)}{\sqrt{\kappa^2+k^2}},\ 
z_{\kappa,k}^{\prime} = \mp\frac{(\kappa+ik)}{\sqrt{\kappa^2+k^2}},
 \end{eqnarray}
 where
 $\mathbf{r}=(R\varphi\cos\theta-\zeta\sin\theta,R\varphi\sin\theta+\zeta\cos\theta)$.

Periodic boundary conditions along the NT circumference
[$\Psi(\mathbf{r+C}_h)=\Psi(\mathbf{r})$] quantize the wave vector associated
with the $\mathbf{C}_h$ direction
[$(\mathbf{k}+\mathbf{K}^{(\prime)})\cdot\mathbf{C}_h=2\pi m,\ m\in\mathbb{Z}$]:
$\kappa\to(m-\tau_3 \nu/3)/R$, where $m\in\mathbb{Z}$ and $\nu=0,\pm1$ is
determined by  $n_1-n_2=3N+\nu$ ($N\in\mathbb{Z}$).\cite{Ando2005}
A NT with $\nu=0$ [e.g. $(n,n)$
armchair NT] has zero band gap and is called a metallic NT. Such a NT is not
suitable to confine particles due to the Klein paradox in gapless structures.\cite{KNG2006} Therefore, semiconducting NTs ($\nu=\pm1$) are more favourable
for quantum dot realizations, and we focus on this case in the following. An
additional feature of semiconducting NTs with $\nu=\pm1$ is that they allow us
to avoid the problem of energy degeneracy at the $\mathbf{K}$ and $\mathbf{K}'$
points by applying an Aharonov -- Bohm flux $\Phi_\mathrm{AB}=B\pi R^2$ through
the NT cross section.\cite{Ando2005} Lifting the degeneracy is crucial for
spin qubit realizations with controlled interqubit exchange.\cite{TBLB2007}
The Aharonov --- Bohm flux leads to a shift of the
quantum number $m\to m+\Phi_\mathrm{AB}/\Phi_0$ ($\Phi_0=hc/|e|$ is
the flux quantum) and to a Zeeman splitting $E_{\kappa_m,k}\to
E_{\kappa_m,k,S_\zeta}=E_{\kappa_m,k}+S_\zeta\hbar\omega_\mathrm{Z}$, where
$\omega_\mathrm{Z}=|e|gB/2m_0c$, $S_\zeta=\pm1/2$ is the spin projection on the
NT axis. 
 Therefore, the energy spectrum and wavefunction of an electron in a NT are given by
\begin{eqnarray}
\label{eq:EnNT}
E_{\kappa_m,k,S_\zeta}&=& E_{\kappa_m,k}+S_\zeta\hbar\omega_\mathrm{Z},\\
\Psi_{\kappa_m,k,S_\zeta}^{(\prime)}(\varphi,\zeta)&=& \Psi_{\kappa_m,k}^{(\prime)}(\varphi,\zeta)|S_\zeta\rangle,
\label{eq:WaveFuncNT}
\end{eqnarray}
where  $\kappa_m =(m+\Phi_\mathrm{AB}/\Phi_0-\tau_3\nu/3)/R$,
$\omega_\mathrm{Z}=|e|gB/2m_0c$, 
 and $|S_\zeta\rangle$ the spin part of the wave function.

\begin{figure}[thb] \begin{center} \includegraphics[width=100mm]{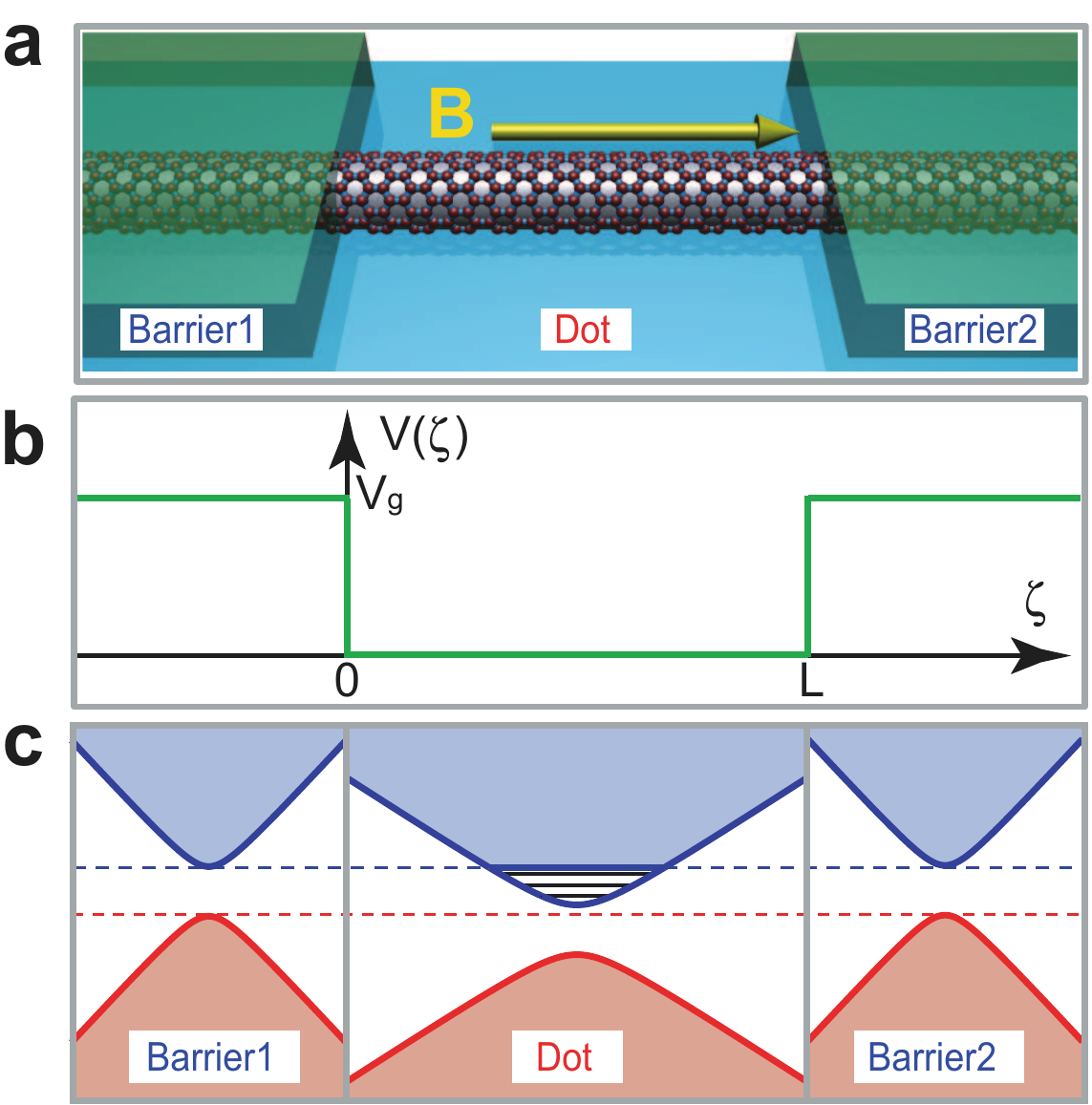}
\end{center}\caption{{ Nanotube quantum dot.} \textbf{a.} Nanotube
    with two top gates. \textbf{b.}
    Longitudinal confinement potential. \textbf{c.} Scheme of the band structure. }    
\label{fig:NT} \end{figure} 

Now we consider a quantum dot (QD) which is made of a NT by the deposition of
top gates on the NT \cite{Mason04,Sapma06,Graeb06} (see Fig.~\ref{fig:NT}a). The spacing between the gates
$L$ defines the length of a QD. We describe the confinement by the
rectangular potential (see Fig.~\ref{fig:NT}b): \begin{equation} V(\zeta)=\left\{
	\begin{array}{cl} V_g, & \zeta<0\ \mathrm{ or }\ \zeta>L,\\ 0,   &
		0\le\zeta\le L.\\ \end{array}\right.  \label{eq:Vg}
	\end{equation}
Recent experimental realizations of a NT QD \cite{Liang2001,Grove,Herrmann} provide clear evidence favouring the rectangular confinement in a QD, since Fabry -- Perot interference observed in such experiments is a testimony for a NT QD with a well-defined length. Note that we consider the experimentally more accessible case, when the length of a NT QD $L$ is much larger than its radius $R$ ($L\approx100\:$nm). For such QDs, the step-like potential drop happens on a length scale much larger than the lattice constant. Therefore, it does not introduce intervalley scattering.

	Straightforward calculations show that the bottom of the
	$m$-th subband of the NT spectrum \textit{under} the top gates $\hbar v
	|\kappa_m|+V_g$ divides the spectrum \textit{between the gates} for this subband into two parts.
	Above the energy $\hbar v |\kappa_m|+V_g$ (dashed blue line in Fig.~\ref{fig:NT}c) the spectrum is continuous
	$E_{\kappa_m,k}$ ($|k|\ge (|V_g|/\hbar v)\sqrt{1+2\hbar v|\kappa_m|/V_g}$) and
	below there is a discrete spectrum $E_{\kappa_m,k_n}$:
\begin{equation}
E_{\kappa_m,k,S_\zeta}=\left\{
\begin{array}{l}
E_{\kappa_m,k_n,S_\zeta},\ (k=k_n\le k_c),\\
E_{\kappa_m,k,S_\zeta},\ (|k|> k_c, k\in\mathbb{R}),\\
\end{array}\right.
\label{eq:Econdis}
\end{equation}
where  $k_c= (|V_g|/\hbar v)\sqrt{1+2\hbar v|\kappa_m|/V_g}$
and allowed values of the quantized wave vector $k_n$ along the NT axis are found from the transcendental equation 
\begin{equation}
\tan k_nL=\frac{(\hbar v)^2\tilde{k}_n k_n}{E_{\kappa_m,k_n}(E_{\kappa_m,k_n}-V_g)-(\hbar v)^2\kappa_m^2}.
\label{eq:kn}
\end{equation}
Here,  $\tilde{k}_n=\sqrt{\kappa_m^2-(E_{\kappa_m,k_n}-V_g)^2/(\hbar v)^2}$. The wavefunction of an electron in a NT QD can be written as follows
\begin{equation}
	\Psi_{\kappa_m,k,S_\zeta}^{(\prime)}(\varphi,\zeta)=\frac{e^{i \mathbf{K^{(\prime)}}\cdot\mathbf{r}}}{\sqrt{2\pi}}e^{i(m-\tau_3\nu/3+\Phi_\mathrm{AB}/\Phi_0)\varphi}
\Phi_{m,k}(\zeta)|S_\zeta\rangle, 
\label{eq:PsiDot}
\end{equation}
where 
%$|S_\zeta\rangle$ is the spin part of the wave function and
\begin{equation}
\Phi_{m,k}(\zeta)=
\left\{\begin{array}{cl}\Phi_{m,k}^L(\zeta),&\zeta<0,\\
\Phi_{m,k}^D(\zeta),&0\le \zeta\le L,\\
\Phi_{m,k}^R(\zeta),& \zeta>L.\\\end{array}\right.
\label{eq:PhiZZ}
\end{equation}
Here, for a discrete spectrum $(k=k_n\le k_c)$:
\begin{eqnarray}
\Phi_{m,k_n}^L(\zeta)&=& A  e^{\tilde{k}_n\zeta}\left( \begin{array}{c}{z}^{(\prime)}_{\kappa_m,-i\tilde{k}_n}\\1\\\end{array} \right),\\
\Phi_{m,k_n}^D(\zeta)&=& \left[C e^{i k_n\zeta}\left( \begin{array}{c}z^{(\prime)}_{\kappa_m,k_n}\\1\\\end{array} \right)+De^{-ik_n\zeta}\left( \begin{array}{c}z^{(\prime)}_{\kappa_m,-k_n}\\1\\\end{array} \right)               \right],\\
\Phi_{m,k_n}^R(\zeta)&=&  B  e^{\tilde{k}_n(L-\zeta)}\left( \begin{array}{c}\
{z}^{(\prime)}_{\kappa_m,i\tilde{k}_n}\\1\\\end{array} \right),
\label{eq:PsiLDR}
\end{eqnarray}
where
\begin{eqnarray}
C&=& A\frac{z^{(\prime)}_{\kappa_m,-k_n}-{z}^{(\prime)}_{\kappa_m,-i\tilde{k}_n}}{z^{(\prime)}_{\kappa_m,-k_n}-z^{(\prime)}_{\kappa_m,k_n}},\ 
D= A\frac{{z}^{(\prime)}_{\kappa_m,-i\tilde{k}_n}-z^{(\prime)}_{\kappa_m,k_n}}{z^{(\prime)}_{\kappa_m,-k_n}-z^{(\prime)}_{\kappa_m,k_n}},\\
B&=&A\left\{  e^{ik_nL}\frac{z^{(\prime)}_{\kappa_m,-k_n}-{z}^{(\prime)}_{\kappa_m,-i\tilde{k}_n}}{z^{(\prime)}_{\kappa_m,-k_n}-z^{(\prime)}_{\kappa_m,k_n}}+ e^{-ik_nL}\frac{{z}^{(\prime)}_{\kappa_m,-i\tilde{k}_n}-z^{(\prime)}_{\kappa_m,k_n}}{z^{(\prime)}_{\kappa_m,-k_n}-z^{(\prime)}_{\kappa_m,k_n}} \right\},
\label{eq:ABCD}
\end{eqnarray}
and $A$ can be found from the normalization condition
\begin{eqnarray}%\nonumber
1&=& \parallel\Psi\parallel^2=|A|^2 [
        ({z}^{(\prime)}_{\kappa_m,-i\tilde{k}_n})^2+1]\frac{1}{2\tilde{k}_n}+|B|^2 [ ({z}^{(\prime)}_{\kappa_m,i\tilde{k}_n})^2+1]\frac{1}{2\tilde{k}_n}+4 L|C|^2 \\ \nonumber
		&&+\mathrm{Re}\left[CD^*[(z^{(\prime)}_{\kappa_m,-k_n})^2+1]
\frac{1}{ik_n}(1-e^{-2ik_nL})\right].
\label{eq:|A|}
\end{eqnarray}
For the $\mathbf{K}$-point, we obtain
\begin{eqnarray}
C&=&  A\frac{1}{2}+i\mathrm{Im}C,\ 
D= A\frac{1}{2}-i\mathrm{Im}C,\\
\mathrm{Im}C&=&\frac{A\kappa_m}{2k_n}\left( -1+\frac{E_{\kappa_m,k_n}}{E_{\kappa_m,k_n}-V_g}\frac{\kappa_m-\tilde{k}_n}{\kappa_m} \right)=-\frac{AE_{\kappa_m,k_n}}{2k_n}\left( \frac{\kappa_m}{E_{\kappa_m,k_n}}-\frac{\kappa_m-\tilde{k}_n}{E_{\kappa_m,k_n}-V_g} \right),\\ 
B&=& A\cos (k_nL)\frac{\kappa_m}{\kappa_m+\tilde{k}_n}\left[1+\frac{\tilde{k}_n^2}{E_{\kappa_m,k_n}(E_{\kappa_m,k_n}-V_g)-(\hbar v \kappa_m)^2} \right].
\label{eq:ABCDKopint}
\end{eqnarray}

For a continuous spectrum $(|k|>k_c)$, we make the following ansatz:
\begin{eqnarray}
\Phi_{m,k}^L(\zeta)&=&  e^{i\tilde{k} \zeta}\left( \begin{array}{c}z_{\kappa_m,\tilde{k}}\\1\\\end{array} \right)+Re^{-i\tilde{k} \zeta}\left( \begin{array}{c}z_{\kappa_m,-\tilde{k}}\\1\\\end{array} \right) ,\\
\label{eq:cL}
\Phi_{m,k}^D(\zeta)&=& A_c e^{ik\zeta}\left( \begin{array}{c}z_{\kappa_m,k}\\1\\\end{array} \right)+B_ce^{-ik \zeta}\left( \begin{array}{c}z_{\kappa_m,-k}\\1\\\end{array} \right),\\
\label{eq:cD}
\Phi_{m,k}^R(\zeta)&=& Te^{i\tilde{k} (\zeta-L)}\left( \begin{array}{c}z_{\kappa_m,\tilde{k}}\\1\\\end{array} \right),
\label{eq:cR}
\end{eqnarray}
where  $|k|\ge(|V_g|/\hbar v)\sqrt{1+2\hbar v |\kappa_m|/V_g}$, $\tilde{k}=\pm\sqrt{[ \left(  E_{\kappa_m,k}-V_g\right)/\hbar v]^2-(\kappa_m)^2}$, and find that
\begin{eqnarray}
\label{eq:Ac}
A_c&=& e^{-ikL}\frac{(z_{\kappa_m,-k}-z_{\kappa_m,\tilde{k}})(z_{\kappa_m,\tilde{k}}-z_{\kappa_m,-\tilde{k}})\sqrt{\kappa_m^2+\tilde{k}^2}\sqrt{\kappa_m^2+k^2}}{4\left[\tilde{k} k \cos (kL) -i\sin(kL)(\sqrt{\kappa_m^2+\tilde{k}^2}\sqrt{\kappa_m^2+k^2}-\kappa_m^2) \right]},\\
B_c&=& \frac{z_{\kappa_m,\tilde{k}}-z_{\kappa_m,-\tilde{k}}}{z_{\kappa_m,-k}-z_{\kappa_m,-\tilde{k}}}-A_c\frac{z_{\kappa_m,k}-z_{\kappa_m,-\tilde{k}}}{z_{\kappa_m,-k}-z_{\kappa_m,-\tilde{k}}},\\
R&=& A_c+B_c-1,\ T= A_ce^{ikL}+B_ce^{-ikL}.
\end{eqnarray}

\section{Spin-orbit interaction in nanotubes}
\label{sec:SOI}

Next, we take spin-orbit interaction (SOI) effects into account. In graphene,
there are two main mechanisms of SOI: Intrinsic SOI,
$H_\mathrm{SO}^\mathrm{int} = \Delta_\mathrm{int} \tau_3\sigma_3 s_z $,
\cite{Kane2005} and extrinsic SOI (Bychkov -- Rashba like), $H_\mathrm{SO}^\mathrm{ext} =
(\Delta_{E}+\Delta_\mathrm{curv})(\tau_3\sigma_1s_y-\sigma_2s_x)$ which is due
to the asymmetric confinement potential normal to the graphene sheet
($\Delta_E$) \cite{Kane2005} and curvature induced effective electric field of
rippled graphene ($\Delta_\mathrm{curv}$) \cite{Guinea2006}  ($s_j$
is the Pauli spin matrix). In a NT, i.e. a graphene sheet rolled up into a
cylinder, the spin components perpendicular to the NT axis become
dependent on the polar angle $\varphi$:\cite{Ando2000}
\begin{eqnarray}
s_x&=&i\left(-S_+e^{i\varphi}+S_-e^{-i\varphi}\right),\\ 
s_y&=&2S_\zeta,\\
s_z&=&S_+e^{i\varphi}+S_-e^{-i\varphi},
\label{eq:sxsysz}
\end{eqnarray}
where, in the eigenbasis of $S_\zeta$, $2S_\zeta|\uparrow\rangle=|\uparrow\rangle$,
$2S_\zeta|\downarrow\rangle=-|\downarrow\rangle$,
$S_+|\uparrow\rangle=S_-|\downarrow\rangle=0$,
$S_+|\downarrow\rangle=|\uparrow\rangle$, and
$S_-|\uparrow\rangle=|\downarrow\rangle$. Therefore, for a NT, the intrinsic
SOI Hamiltonian is given by
\begin{equation}
H_\mathrm{SO}^\mathrm{int}=\Delta_\mathrm{int} \tau_3\sigma_3(S_+e^{i\varphi}+S_-e^{-i\varphi}),
\label{eq:HSO_int}
\end{equation}
the extrinsic SOI term due to $\Delta_E$ is given by
\begin{eqnarray}
%\nonumber
H_\mathrm{SO}^E&=&  \Delta_E\left[2\tau_3\sigma_1S_\zeta -i\sigma_2\left(-S_+e^{i\varphi}+S_-e^{-i\varphi}\right)  \right],
\label{eq:HSO_E}
\end{eqnarray}
and the extrinsic SOI term due to curvature of a NT is given by \cite{Ando2000}
\begin{equation}
H_\mathrm{SO}^\mathrm{curv}=i\Delta_\mathrm{curv}^\perp\sigma_2 \left(-S_+e^{i\varphi}+S_-e^{-i\varphi}\right)+\Delta_\mathrm{curv}^\parallel \tau_3\sigma_12S_\zeta,
\label{eq:HSO_curv}
\end{equation}
where
$\Delta_\mathrm{curv}^\perp=-\Delta(V_{pp}^\sigma-V_{pp}^\pi)a_0/8\sqrt{3}R\varepsilon_{\pi\sigma}$
and
$\Delta_\mathrm{curv}^\parallel=\Delta(3V_{pp}^\sigma+5V_{pp}^\pi)a_0/8\sqrt{3}R\varepsilon_{\pi\sigma}$
($\Delta=12\:$meV,\cite{Serrano2000} $\varepsilon_{\pi\sigma}=7.3\:$eV,
$V_{pp}^\sigma=6.38\:$eV, and $V_{pp}^\pi=-2.66\:$eV [\onlinecite{Tomanek1988}]).
Note that at moderate electric fields ($E<0.1\:$V$/$nm), the last SOI term is
dominant ($\Delta_\mathrm{int}\approx1\:\mu$eV,\cite{Guinea2006}
$\Delta_E<\Delta_\mathrm{int}$, and
$\Delta_\mathrm{curv}^\perp\approx-(0.26\:\mathrm{meV}/R[\mathrm{nm}])$) and,
therefore, the other types of SOI can be safely neglected.

\begin{figure}[thb]   \begin{center} \includegraphics[width=100mm]{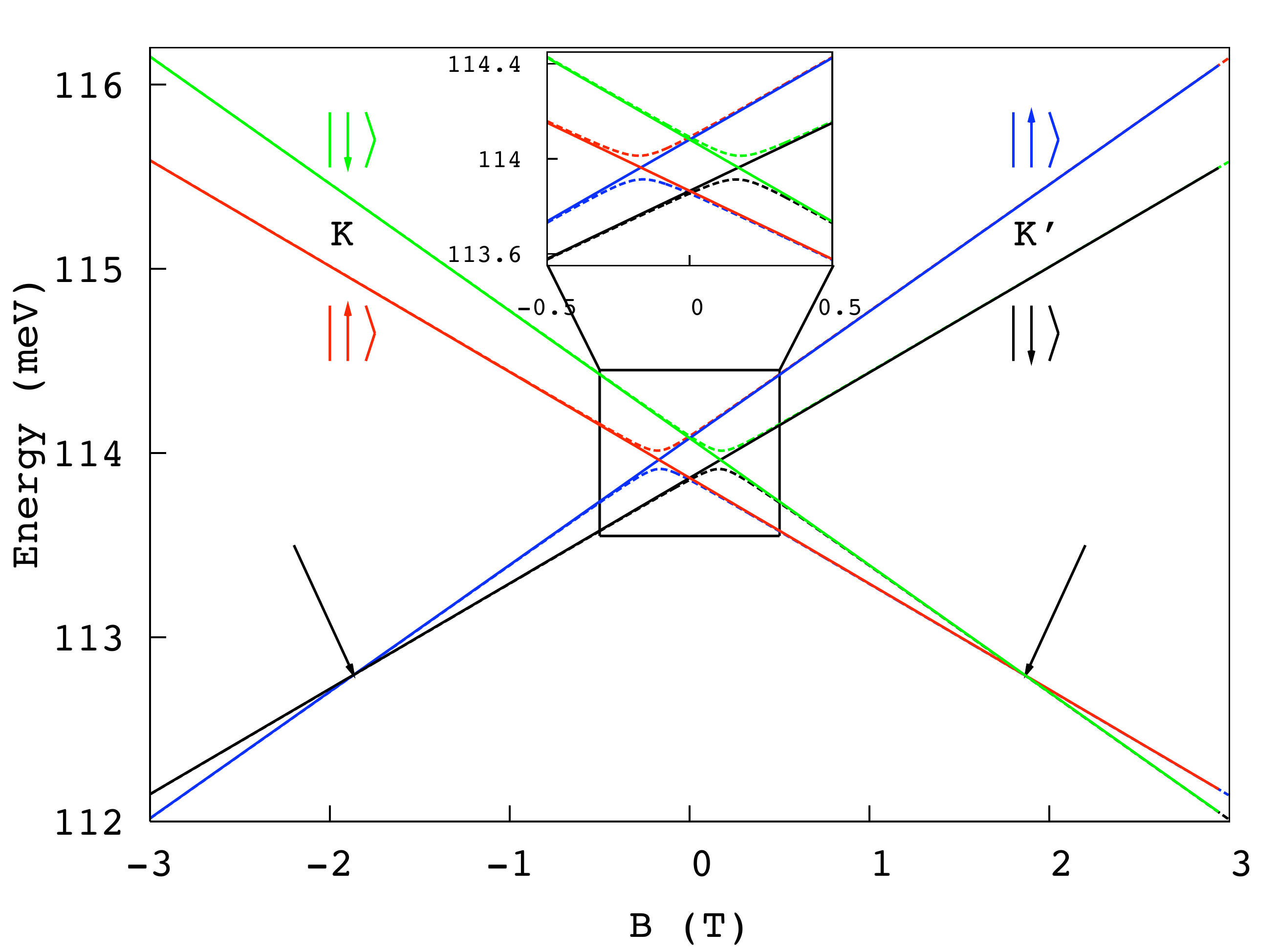}
\end{center} 
\caption{ Lowest energy levels of electrons in a NT QD at low
magnetic fields ($R\approx1.6\:$nm, $L=100\:$nm, $g=2$, and $V_g=\hbar v /40
R\approx8.5\:$meV). Solid curves correspond to the case of zero intervalley
mixing. At zero magnetic fields there is splitting of the levels due to the
second term in Eq.~(\ref{eq:HSO_curv}). The magnitude of the splitting is
$2\Delta_\mathrm{curv}^\parallel=0.22\:$meV. The arrows indicate crossings of
$|\uparrow\rangle$ and  $|\downarrow\rangle$ states of a certain $\mathbf{K}$
point. At these levelcrossings, a singularity appears in $1/T_1$ (see
below in Fig.~\ref{fig:One}). Dashed curves correspond to the case of weak
intervalley mixing ($\Delta_\mathbf{K-K'}=0.05\:$meV). It modifies the
zero-field splitting $2|\Delta_\mathrm{curv}^\parallel|\to2\sqrt{
  {\Delta_\mathrm{curv}^\parallel}^2+\Delta_\mathbf{K-K'}^2}$ and opens up avoided crossings
(with the value $2|\Delta_\mathbf{K-K'}|$)  of the levels with the same spin
orientation but different valley index. This is illustrated in the inset which
is a blow-up of the center region of the spectrum.}  
\label{fig:spectrum} \end{figure}

%--+--
	The last term  
$\propto\Delta_\mathrm{curv}^\parallel S_\zeta$ (where
$\Delta_\mathrm{curv}^\parallel\approx0.17\:\mathrm{meV}/R[\mathrm{nm}]$) in
Eq.~(\ref{eq:HSO_curv}) leads to a shift 
$\kappa_m\to\varkappa_m^\pm=\kappa_m\pm\Delta^\parallel_{\mathrm{curv}}/\hbar
v$ \cite{Ando2000} (where $\pm$ corresponds here to $|\uparrow \rangle$ and
$|\downarrow \rangle$ states)  and, therefore, to a spin splitting:
\begin{equation}
		E_{\varkappa_m^+,k_n,+1/2}-E_{\varkappa_m^-,k_n,-1/2}\approx\hbar\omega_\mathrm{Z}-2\,\mathrm{sgn}(m-\tau_3\nu/3)\Delta_\mathrm{curv}^\parallel
\end{equation}
	(for $|\varkappa_m^\pm|\gg k_n$).  Thus, SOI 
	$\propto\Delta_\mathrm{curv}^\parallel$ acts as an effective magnetic field resulting
	in a level splitting ($2\Delta_\mathrm{curv}^\parallel$) at zero magnetic field,
	as has been now experimentally confirmed now.\cite{McEuen2007}
	Note that this zero-field splitting does not violate Kramers theorem,
	since time reversed states correspond to different non-equivalent
	$\mathbf{K}$-points and are degenerate at zero $B$-fields (see Fig.~\ref{fig:spectrum}). The existence of the zero-field splitting opens up
	an intriguing possibility for spin resonance experiments
	\textit{without} any magnetic fields: the first term in
	Eq.~(\ref{eq:HSO_curv}) allows electric-dipole transitions between
	spin-up and spin-down states, the second term (as an effective magnetic
	field) splits these states, and thus oscillating electric fields
	perpendicular to a NT lead to electric-dipole spin resonance with
	resonance frequency $\omega=2\Delta_\mathrm{curv}^\parallel/\hbar\approx
	33\times10^{10}\:$s$^{-1}$ and Rabi frequency $\omega_\mathrm{R}\approx
	1.6\times10^5\:$s$^{-1}$ at $E=10\:$V$/\:$cm and $V_g=2.3\:$meV (see \appendixname~\ref{EDSR}).

	We know that intervalley mixing leads to splitting
        of the levels corresponding to different $\mathbf{K}$ points which has
        been observed in Ref.~\onlinecite{JH2005}. Such
        mixing does not split Kramers doublets (in the case of time reversal
        symmetric intervalley scattering) but modifies the magnitude of the
        splitting ($2\Delta_\mathrm{curv}^\parallel\to2\sqrt{
          {\Delta_\mathrm{curv}^\parallel}^2+\Delta_\mathbf{K-K'}^2}$, where
        $\Delta_\mathbf{K-K'}$ is the intervalley mixing strength) between
        spin-up and spin-down states of a certain $\mathbf{K}$-point and leads
        to anticrossings at non-zero $B$-fields of the levels with the same
        spin orientation but belonging to different $\mathbf{K}$ valleys (see
	Fig.~\ref{fig:spectrum} and \appendixname~\ref{valley}). 
	
	In the case of negative voltage applied to the top gates ($V_g<0$),
        hole states become localized instead of electrons. It can be shown
        that the energy spectrum of the lowest levels of holes has the same
        structure as for electrons (illustrated in Fig.~\ref{fig:spectrum})
        but shifted down by the energy gap $E_g\approx2\hbar
        v\sqrt{\kappa_0^2+k_0^2}\approx228\:$meV. 
	From Fig.~\ref{fig:spectrum} we see that electron energy levels cross
        at $\hbar\omega_\mathrm{Z}=2\tau_3\nu\Delta_\mathrm{curv}^\parallel$ (indicated by
        arrows in Fig.~\ref{fig:spectrum}), whereas there are no crossings of
        the two highest levels of holes at nonzero magnetic fields. Therefore, SOI
        (due to zero-field splitting of energy levels) breaks the
        electron-hole symmetry. For the estimation of the
        SOI constants we use band parameters of bulk graphite. Note that for small
        radius NT, due to curvature effects, strong hybridization of bands can
        modify the band paramenters of a NT and, thus, the SOI constants or
        the $g$-factor. If the SOI constant had the
        opposite sign due to hybridization, then the energy spectrum for
		electrons would look like the one for holes and vice versa \cite{McEuen2007}. Hence, in
        the case of negative $\Delta_\mathrm{curv}^\parallel$, there are
        crossings of levels for holes and not for electrons (at $B\ne0$). Such
        electron-hole asymmetry in the spectrum can provide us with
        information about the sign and the magnitude of the SOI constant and
        about the $g$-factor.

Now, we turn to the quantitative discussion of the spin relaxation time 
in nanotube quantum dots.
We take the first term in Eq.~(\ref{eq:HSO_curv}) into
account in the framework of perturbation theory, 
which leads to the solution of the Dirac (eigenvalue) equation for the lowest levels
$(H_0\pm\hbar\omega_\mathrm{Z}/2+
H_\mathrm{SO}^\mathrm{curv})\psi_{0,0,\pm1/2}=\mathcal{E}_{0,0,\pm1/2}\psi_{0,0,\pm1/2}$
in first order in $H_\mathrm{SO}^\mathrm{curv}$:
\begin{eqnarray}
\label{eq:energy}
\mathcal{E}_{0,0,\pm1/2}&\approx& E_{\varkappa_0^\pm,k_0,\pm1/2},\\
\label{eq:psi}
\psi_{0,0,\pm1/2}(\varphi,\zeta)&\approx& \Psi_{\varkappa_0^\pm,k_0,\pm1/2}(\varphi,\zeta)
+\sum_{n\neq0}\lambda_{k_n}^\mp\Psi_{\varkappa_{\mp1}^\mp,k_n,\mp1/2}(\varphi,\zeta)
+\frac{L}{2\pi}\int_{\pm k_c}^{\pm\infty}dk\lambda_{k}^\mp\Psi_{\varkappa_{\mp1}^\mp,k,\mp1/2}(\varphi,\zeta),\\
\label{eq:lambdamp}
\lambda_{k}^\pm&=& \pm i\Delta_\mathrm{curv}^\perp\frac{\langle \Phi_{\varkappa_{\pm1}^\pm,k}(\zeta)|\sigma_2|\Phi_{\varkappa_0^\mp,k_0}(\zeta)\rangle}{E_{\varkappa_{\pm1}^\pm,k,\pm1/2}-E_{\varkappa_0^\mp,k_0,\mp1/2}}.
\end{eqnarray}
Note that the function $\left(\Phi_{\varkappa_{m'}^\pm,k_{n'}}(\zeta)\right)^\dagger\sigma_2\Phi_{\varkappa_m^\mp,k_n}(\zeta)$ is either symmetric or antisymmetric with respect to inversion at $\zeta=L/2$. Hence, there is a selection rule for SOI between quantized levels, namely,
\begin{equation}
	\langle \Phi_{\varkappa_{m'}^\pm,k_{n'}}(\zeta)|\sigma_2|\Phi_{\varkappa_m^\mp,k_n}(\zeta)\rangle\propto1-\mathrm{sgn}(m'-1/3)\mathrm{sgn}(m-1/3)(-1)^{n'+n}.
	\label{eq:selrules}
\end{equation}
Thus, $\lambda_{k_{n'}}^+=0$ ($\lambda_{k_{n'}}^-=0$) for odd (even) $n'$.

\section{Electron-phonon coupling in nanotubes}
\label{sec:ElPh}

For definiteness, we consider only such ($n_1,n_2$) NTs that
$n_1-n_2=3N+1$ ($\nu=1$).  Then the two states  ($\psi_{0,0,\pm1/2}$) with the lowest energy (at $\Phi_\mathrm{AB}/\Phi_0>\Delta_\mathrm{curv}^\parallel R/\hbar v$) of a NT QD belong to the $\mathbf{K}$ point
with $\tau_3=1$ (see Fig.~\ref{fig:spectrum}). Phonon induced transitions (which become allowed due to SOI)
between these states give the dominant contribution to spin relaxation of a
single particle in a NT QD. Despite quite complicated phonon dispersion
relations in NTs,\cite{Saito1999} it is possible to find analytical
expressions for the electron-phonon coupling in NTs in the case of low-energy
phonons.

 The splitting between the $\mathcal{E}_{0,0,\pm1/2}$ states is less than $1\:$meV at $B<10\:$T.
Phonons with much higher energies are not favourable for transitions between these levels. The energy of the radial breathing mode is $\hbar\omega_\mathrm{RBM}>8.5\:$meV [\onlinecite{Maultzsch2005}]
for NTs with $R\le1.5\:$nm, which excludes that mode (and all higher modes)
from our analysis. Thus, only three acoustic phonon modes are
important for spin-flip transitions between the lowest two levels: the
twisting mode (TM), the stretching mode (SM), and the bending mode (BM).\cite{Ando2005} To describe these modes we use a continuum model
\cite{Ando2005} in which the equation of motion for the displacement
$\mathbf{u}(\mathbf{r},t)=(u_\varphi,u_\zeta,u_r)$ is given by
\begin{equation}
\ddot{\mathbf{u}}(\mathbf{r},t)=\Lambda \mathbf{u}(\mathbf{r},t),
\label{eq:uphonon}
\end{equation}
where the force-constant tensor
\begin{equation}
\Lambda=\left( \begin{array}{ccc}
\frac{c_l^2}{R^2}\nabla^2_{\varphi\varphi}+c_t^2\nabla^2_{\zeta\zeta}&\frac{c_l^2-c_t^2}{R}\nabla^2_{\varphi\zeta}&\frac{c_l^2}{R}\nabla_r\\
\frac{c_l^2-c_t^2}{R}\nabla^2_{\varphi\zeta}&c_l^2\nabla^2_{\zeta\zeta}+\frac{c_t^2}{R^2}\nabla^2_{\varphi\varphi}&\frac{c_l^2-2c_t^2}{R}\nabla_r\\
-\frac{c_l^2}{R}\nabla_r&-\frac{c_l^2-2c_t^2}{R}\nabla_\zeta&-\frac{c_l^2}{R^2}
\end{array} \right)
\label{eq:phonon_matrix}
\end{equation}
is invariant under the group
symmetry operations of a NT.\cite{Ando2005,Goupalov2005} Here, $c_l$ and $c_t$ are the longitudinal and transverse phonon velocities,
respectively ($c_l=20.9\:$km$/$s and $c_t=12.3\:$km$/$s [\onlinecite{Suzuura2002}]). Substituting the
solution of Eq.~(\ref{eq:uphonon}) in the form
$\mathbf{u}(\mathbf{r},t)=\mathbf{A}_\alpha\exp[i(m\varphi+q\zeta-\omega t)]$
($q$ and $\omega$ are the phonon wave vector and frequency, respectively, and
$\alpha$ is the phonon mode) and keeping only leading terms in $qR$
($qR\ll1$), we get for TM phonons ($m=0$):
\begin{equation}
\omega_\mathrm{T}=c_tq,\ \mathbf{A}_\mathrm{T}=A_\mathrm{T}(1,0,0),
\label{eq:TM}
\end{equation}
for SM phonons ($m=0$):
\begin{equation}
\omega_\mathrm{S}= c_\mathrm{S}q,\ \mathbf{A}_\mathrm{S}=A_\mathrm{S}\left(0,1,-iqR\eta\right),
\label{eq:SM}
\end{equation}
and for BM phonons ($m=1$):
\begin{eqnarray}
\label{eq:w=q2}
\omega_\mathrm{B}&=&c_\mathrm{S}Rq^2/\sqrt{2},\\
\mathbf{A}_\mathrm{B}&=& \frac{A_\mathrm{B}}{\sqrt{2}}\left(i+\frac{i\eta (qR)^2}{2},-iqR,1-\frac{\eta (qR)^2}{2}  \right),
\label{eq:B-mode}
\end{eqnarray}
where  $c_\mathrm{S}= 2(c_t/c_l)\sqrt{c_l^2-c_t^2}$,
$\eta=(c_l^2-2c_t^2)/c_l^2$; $A_j=\sqrt{\hbar/2M\omega_j}$ ($M$ is the NT
mass). We see that TM and SM show linear dispersion, whereas BM exhibits quadratic dispersion. Note that these results are only valid for
long-wavelength phonons ($qR\ll1$ and $\omega<\omega_\mathrm{RBM}$).

The electron-phonon coupling is expressed by the operator
\begin{equation}
V_\mathrm{el-ph}=\left( \begin{array}{cc}
V_1&V_2\\
V_2^*&V_1
\end{array} \right) + \mathrm{H.c.},
\label{eq:H_el-ph}
\end{equation}
where for the $\mathbf{K}$-point 
\begin{eqnarray}
V_1&=& g_1(u_{\varphi\varphi}+u_{\zeta\zeta}),\ V_2=g_2e^{3i\theta}(u_{\varphi\varphi}-u_{\zeta\zeta}+2iu_{\varphi\zeta}),\\
u_{\varphi\varphi}&=&\frac1R \frac{\partial u_\varphi}{\partial \varphi}+\frac{u_r}{R},\ u_{\zeta\zeta}=\frac{\partial u_\zeta}{\partial \zeta},\ 2u_{\varphi\zeta}=\frac{\partial u_\varphi}{\partial \zeta}+\frac1R\frac{\partial u_\zeta}{\partial \varphi},
\label{eq:V1&V2}
\end{eqnarray}
$g_1\approx30\:$eV is the deformation potential constant (which appears in diagonal elements of $V_\mathrm{el-ph}$), and the off-diagonal coupling constant $g_2\approx1.5\:$eV (which is caused by change in the bond-length between neighboring carbon atoms). \cite{Suzuura2002}   Using Eqs.~(\ref{eq:TM})--(\ref{eq:H_el-ph}), we get for the TM:
\begin{eqnarray}
V_1^\mathrm{T}&=& 0,\ V_2^\mathrm{T}= -g_2A_\mathrm{T}qe^{3i\theta}e^{i(q\zeta-\omega_\mathrm{T} t)},
\label{eq:T}
\end{eqnarray}
for the SM:
\begin{equation}
V_1^\mathrm{S}= 2 i g_1A_\mathrm{S} q \frac{c_t^2}{c_l^2}e^{i(q\zeta-\omega_\mathrm{S} t)},\ V_2^\mathrm{S}= i g_2A_\mathrm{S} q \frac{c_\mathrm{S}^2}{c_t^2}e^{3i\theta}e^{i(q\zeta-\omega_\mathrm{S} t)},
\label{eq:S}
\end{equation}
and for the BM:
\begin{eqnarray}V_1^\mathrm{B}&=&\sqrt{2} g_1A_\mathrm{B}  q^2 R \frac{c_t^2}{c_l^2}e^{i(\varphi+ q\zeta-\omega_\mathrm{B} t)},\ 
V_2^\mathrm{B}= g_2A_\mathrm{B}  q^2R \frac{c_\mathrm{S}^2}{\sqrt{2}c_t^2}e^{3i\theta}e^{i(\varphi+q\zeta-\omega_\mathrm{B} t)}.
\label{eq:B}
\end{eqnarray}

%{\it \textbf{May be we delete this:} It can be seen that the electron-phonon coupling potential for the stretching and twisting modes (the bending mode) is proportional to $q$ ($q^2$), therefore, the energy of phonon modes ($\langle \psi_1|V_\mathrm{el-ph}|\psi_2\rangle$) is inversely proportional to $L$ ($L^2$), as was obtained in an experiment \cite{Sapmaz2005}. In other words, for the twisting and stretching modes, $\omega\propto q$ (for the bending mode, $\omega\propto q^2$), thus the phonon energy $\hbar\omega\propto L^{-1}$ ($\hbar\omega\propto L^{-2}$), as long as $q\propto L^{-1}$.}

%--+--
Note that the electron-phonon coupling in nanotubes is very 
strong (for example, compare $g_1\approx30\:$eV with a deformational acoustic coupling constant in GaAs $\Xi_0\approx6.5\:$eV). Furthermore, the electron wave function is highly localized in the dot region (it decays 
exponentially outside the dot). Thus, the phonons in the contacts or substrate can be safely ignored for our purposes. Moreover, we neglect the effect of the substrate on the phonon modes. This
is justified due to the
relatively weak coupling between the substrate and the NT, very high
stiffness 
and rigidity of a NT, and, last but not least, very small atomic
displacement amplitudes  in an acoustic phonon wave (which is a few percents
of Angstroms only).

\section{Spin relaxation in nanotubes}
\label{sec:T1}

We are now able to analyze spin-flip transitions between the lowest energy
levels induced by long-wavelength phonons. Using Eq.~(\ref{eq:psi}), the
matrix element of such a transition is given by 
\begin{eqnarray}
	M_{\omega}&\equiv&\nonumber
	\langle\psi_{0,0,-1/2}|V_\mathrm{el-ph}|\psi_{0,0,1/2}\rangle\\ &=&
	\sum_k\left\{ \lambda_{k}^-\langle
	\Psi_{\varkappa_0^-,k_0,-1/2}|V_\mathrm{el-ph}|\Psi_{\varkappa_{-1}^-,k,-1/2}\rangle
	%\right.\\
	%&&\left.
	+(\lambda_{k}^+)^*\langle
	\Psi_{\varkappa_1^+,k,1/2}|V_\mathrm{el-ph}|\Psi_{\varkappa_0^+,k_0,1/2}\rangle
	\right\}.\label{eq:M_wB}
\end{eqnarray} 
Here the sum stands for summation over the discrete $k_n$ and integration over the continuous $|k|\ge (|V_g|/\hbar v)\sqrt{1+2\hbar v|\kappa_m|/V_g}$.

From Eq.~(\ref{eq:PsiDot}),
$\langle\Psi_{\kappa_{m_1},k_n,\pm1/2}|e^{im_2\varphi}e^{iq\zeta}|\Psi_{\kappa_{m_3},k_{n'},\pm1/2}\rangle=\langle\Phi_{m_1,k_n}|e^{iq\zeta}|\Phi_{m_3,k_{n'}}\rangle
\delta_{m_1,m_2+m_3}$. Therefore, only phonon modes with $m_2=1$ give non-zero
contribution to spin-flip transitions (this is an additional reason why we do not need to consider
higher phonon modes with $m_2>1$). 
Thus, only BM-phonons are responsible for the spin
relaxation, whereas TM- and SM-phonons (with $m_2=0$) cannot flip the spin.

In the framework of Bloch -- Redfield theory,\cite{Blum} the spin
relaxation time induced by BM-phonons is given by 
\begin{eqnarray}\nonumber
\frac{1}{T_1}&=&\frac{2\pi}{\hbar}L\int_{-\infty}^\infty dq(2N_\omega+1)|M_\omega|^2\delta\left( \hbar\omega_0-\hbar\frac{c_\mathrm{S}R}{\sqrt{2}q^2} \right)\\
\nonumber
&=& \frac{2\pi}{\hbar}L\int_{-\infty}^\infty dq(2N_\omega+1)|M_\omega|^2\frac{1}{2^{3/4}\hbar\sqrt{c_\mathrm{S}R\omega_0}}\left[ \delta(q-q_0)+\delta(q+q_0) \right]\\
\label{eq:T1}
&=&   \frac{2^{5/4}\pi
L}{\hbar^2\sqrt{c_\mathrm{S}R\omega_0}}
\left(2N_{\omega_0}+1 \right)
\left|M_{\omega_0}\right|^2, 
\end{eqnarray} 
where
$\omega_0=|\mathcal{E}_{\varkappa_0^+,k_0,+1/2}-\mathcal{E}_{\varkappa_0^-,k_0,-1/2}|/\hbar\approx|\omega_\mathrm{Z}-2\tau_3\Delta_\mathrm{curv}^\parallel/\hbar|$,  $q_0=\sqrt{\sqrt{2}\omega_0/c_\mathrm{S}R}$, and
$N_\omega=[\exp(\hbar\omega/k_\mathrm{B}T)-1]^{-1}$ is the Bose
distribution function. 
Note that pure dephasing $1/T_\varphi=0$ for BM phonons and
$1/T_\varphi=O\left( \Delta_\mathrm{SO}^4 \right)$ for SM and TM phonons,
therefore,  $1/T_2=1/2T_1+1/T_\varphi=1/2T_1$  in
first-order perturbation theory.   

 We used the Markov and the secular approximations in the derivation of
 Eq.~(\ref{eq:T1}). We can estimate the correlation time in the phonon bath to
 be $\tau_c\approx1\:$ps. Therefore, the Markov approximation ($T_1\gg\tau_c$)
 and the secular approximation ($\omega_0T_1\gg1$) are valid except
 for the energy regime close to the level crossing at $\omega_0=0$.
 Moreover, our estimations of the electron-phonon coupling are valid for
 phonons with the wavelength shorter than the full length of the NT $l_{NT}$.
 Therefore, in the case of a small splitting between spin-up and spin-down
 states (long wavelength phonons), the results are trustworthy for sufficiently
 long NTs ($l_{NT}q_0\gg1$), for example, if the spin splitting is
 $1\mu$eV, then the NT length should be greater than $700\:$nm.

 We now study spin relaxation induced by low-frequency phonons
 ($\omega_0\approx|\omega_\mathrm{Z}-2\tau_3\Delta_\mathrm{curv}^\parallel|\to0$). As
 shown above, such spin relaxation occurs near the level crossing indicated
 by arrows in Fig.~\ref{fig:spectrum}. 
%Therefore, the current discussion
%is relevant for electrons (if $g$-factor and $\Delta_\mathrm{curv}^\parallel$
%have the same signs) or for holes (if those signs are opposite). 
 One can show that $|M_{\omega_0}|^2\propto\omega_0$ and
 $N_{\omega_0}\propto T/\omega_0$ (at
 $k_\mathrm{B}T\gg\hbar\omega_0$) for $\omega_0\to0$.
Moreover, the density of states for one-dimensional phonon modes with
quadratic dispersion, i.e. the bending modes responsible for spin relaxation,
has a van Hove singularity at zero frequency. It goes like 
$1/\sqrt{\omega_0}$ where $\omega_0$ is the phonon frequency of the bending mode. 
This translates into the existence of a singularity in
   the noise spectral function $J(\omega_0)\propto1/\sqrt{\omega_0}$ which 
describes particle spin relaxation due to coupling to NT
lattice vibrations via SOI and electron-phonon interaction. Therefore,
\begin{equation}
	\label{eq:rose_noise}
1/T_1\propto1/\sqrt{\omega_0}
\end{equation} 
at low $\omega_0$. To the best of
our knowledge, this is the first system that exhibits a $1/\sqrt{\omega_0}$ 
spin-phonon
noise spectrum at low frequencies. 
 Such a result (fast relaxation times at small
 splitting between spin-up and spin-down levels) is counter-intuitive in the
 light of the commonly expected long $T_1$ time for NTs (due to the expected weak
 SOI) and  compared to the usual behaviour of the spin relaxation time
 ($1/T_1\propto\omega_\mathrm{Z}^4$ at low magnetic fields) for conventional
 GaAs QDs.\cite{GKL}

\begin{figure}[thb]
\begin{center}
\includegraphics[width=10cm]{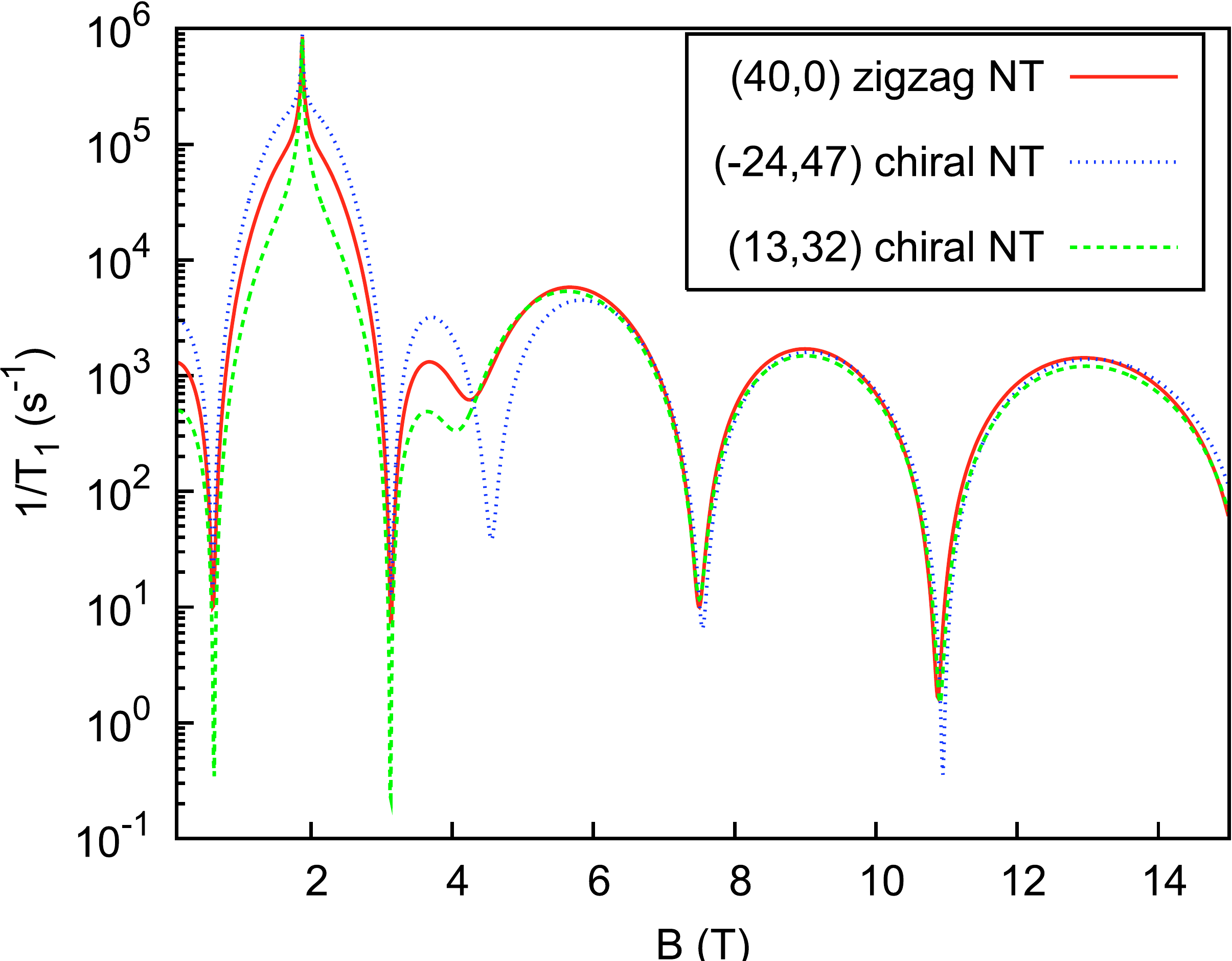}
\end{center}
\caption{{ Chirality dependence of the electron spin relaxation for NT QDs.} The
  spin relaxation rate as a function of a field for a (40,0) zigzag NT (solid
  curve, $\theta=0$), (-24,47) NT (dotted curve, $\theta\approx\pi/2$), and
  (13,32) NT (dashed curve, $\theta\approx \pi/4$). $R\approx1.6\:$nm,
  $L=100\:$nm, $g=2$, $V_g=\hbar v /40R\approx8.5\:$meV, $T=0.1\:$K.} 
\label{fig:Two}
\end{figure}

\begin{figure}[thb]
\begin{center}
\includegraphics[width=10cm]{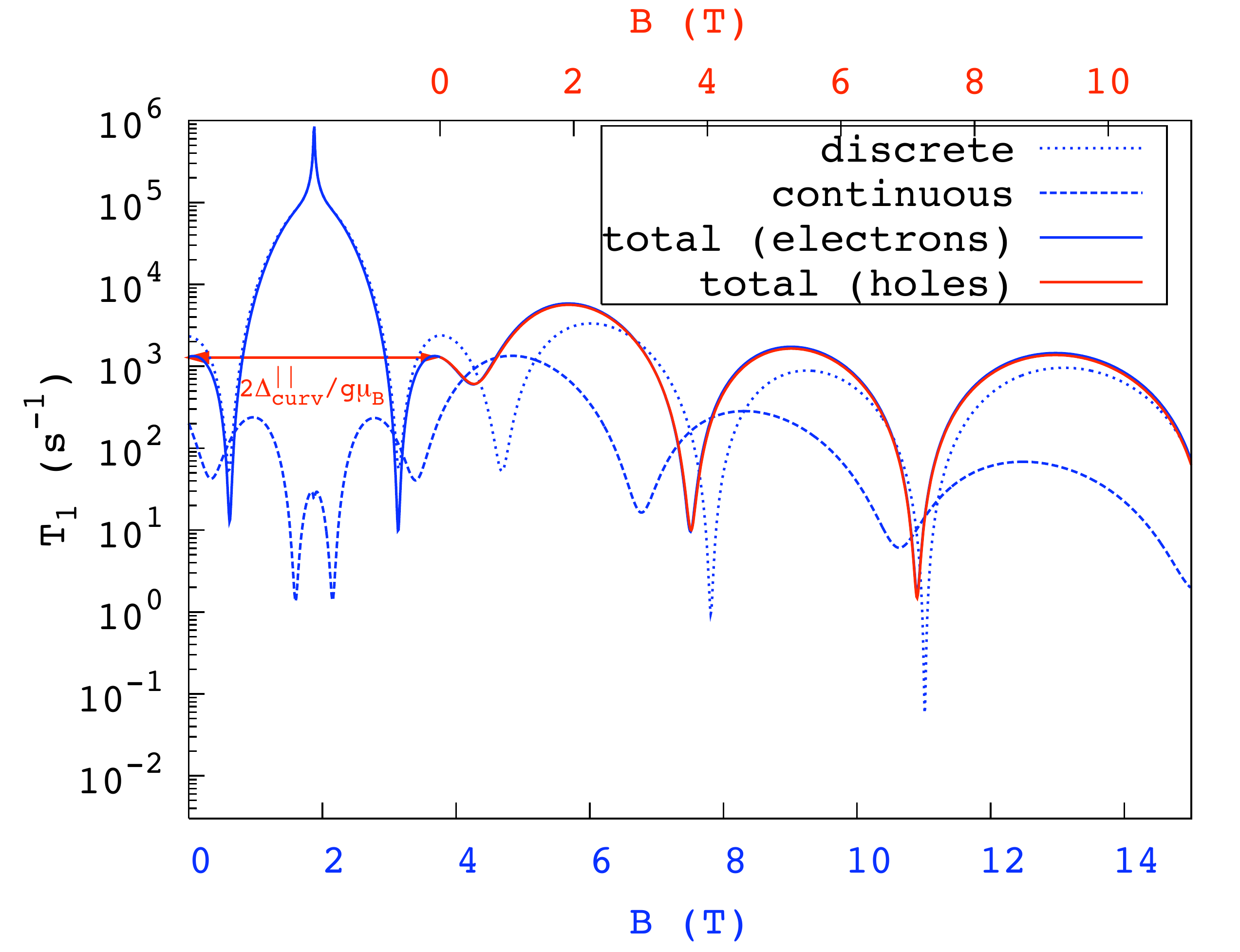}
\end{center}
\caption{ { Spin relaxation of a (40,0) zigzag NT.} The dependence of the electron and hole
  spin relaxation rates $1/T_1$ on a parallel magnetic field $B$ (lower horizontal axis for electrons and upper one for holes) due to SO
  coupling of the two lowest states ($\Psi_{\varkappa_0^\pm,k_0,\pm1/2}$) to
  higher states of the discrete spectrum
  ($\Psi_{\varkappa_{\mp1}^\mp,k_n,\pm1/2}$) is shown by the dashed curve and that
  due to SO coupling to states of the continuous spectrum
  ($\Psi_{\varkappa_{\mp1}\mp,k,\mp1/2}$) by the dotted curve. The total
  spin relaxation rate (solid curves) exhibits destructive interference at each
  odd crossing of the curves related to the abovementioned contributions
  ($R\approx1.6\:$nm, $L=100\:$nm, $g=2$, $T=0.1\:$K,  $V_g=\pm\hbar v
  /40R\approx\pm8.5\:$meV, where the upper/lower sign is for electrons/holes). The electron (blue solid curve) and the hole (red solid curve) spin relaxation time map onto each other by a shift along the magnetic field axis by $\Delta B=2\Delta_\mathrm{curv}^\parallel/g\mu_\mathrm{B}$.} 
\label{fig:Three}
\end{figure}

To better understand Eq.~(\ref{eq:rose_noise}), 
we consider the spectral density of
 the electron-phonon correlation function
 \[J_{mk}(\omega)=\int_{-\infty}^\infty dt\overline{\langle m|
   V_\mathrm{el-ph}(0)|k\rangle\langle k|V_\mathrm{el-ph}(t)|m\rangle}
 e^{i\omega t},
 \]
where the overbar denotes the ensemble average. We first analyze this expression for GaAs QDs and later on for NT QDs.
 
 For the phonon-induced relaxation rate between levels split
 by the Zeeman term, we find $1/T_1\propto
 J_{12}(\omega_\mathrm{Z})\propto\sum_{\mathbf{q}}\sum_l(N_{\omega}+1/2)|\langle1|A_\omega e^{i\mathbf{qr}}|l\rangle\langle l|H_{\mathrm{SO}}^\mathrm{curv}|2\rangle|^2\delta(\omega-\omega_\mathrm{Z})$,   
where $A_\omega$ is the electron-phonon coupling strength and $\omega$ is the
phonon frequency. Therefore, the corellation function defines the phonon-induced electron spin relaxation times. Let $d$ be the single phonon degree of freedom (related to
the dimensionality of the underlying lattice structure). Then, for
GaAs semiconductor structures with linear in momentum $H^\mathrm{curv}_\mathrm{SO}$, we get 
$\sum_\mathbf{q}\to\int dq q^{d-1}\int d\Omega_\mathbf{q}$, 
$\langle1|e^{i\mathbf{qr}}|l\rangle\propto q$ (in dipole approximation),
$\langle l|H^\mathrm{curv}_{\mathrm{SO}}|2\rangle\propto \omega_\mathrm{Z}$, 
$N_{\omega}\propto1/\omega$ (at $k_\mathrm{B}T\gg\hbar\omega$). 
Taking into account that $A_\omega\propto1/\sqrt{\omega}$
 for the coupling between an electron  and a piezoelectric phonon, we obtain  $J_{12}(\omega_\mathrm{Z})\propto \omega_\mathrm{Z}^{d+1}$
  in the case of linear dispersion of a phonon ($\omega\propto q$).
 For deformational acoustic phonons,  $A_\omega\propto\sqrt{\omega}$, therefore, 
$J_{12}(\omega_\mathrm{Z})\propto \omega_\mathrm{Z}^{d+3}$.
 Therefore, at low
 frequency,  the spectral density function of the electron-phonon coupling is
 super-Ohmic ($J(\omega)\propto \omega^n$ $n>1$) even
 for all phonons in all dimensions.

This is fundamentally different for the NT QDs discussed here: 
Since $H^\mathrm{curv}_\mathrm{SO}$ in a NT couples spin to the azimuthal
 degree of freedom (see Eq.~(\ref{eq:HSO_curv})) and the azimuthal component
 of the phonon wave vector is quantized 
(see Eq.~(\ref{eq:B})), we get $\langle1|A_\omega e^{i\mathbf{qr}}|l\rangle\langle
l|H^\mathrm{curv}_{\mathrm{SO}}|2\rangle\propto1+O(q)$. Thus, for deformation-acoustic
phonons ($A_\omega\propto\sqrt{\omega}$) with quadratic dispersion
($\sum_\mathbf{q}\to\int d\omega/\sqrt{\omega}$), we obtain
$J_{12}(\omega_0)\propto 1/\sqrt{\omega_0}$ and recover
Eq.~(\ref{eq:rose_noise}). The noise spectral function $J_{12}(\omega)$ 
describes particle spin dissipation due to coupling to NT
lattice vibrations (via SOI and electron-phonon interaction). 

\begin{figure*}[thb]
	\begin{center} \includegraphics[width=160mm]{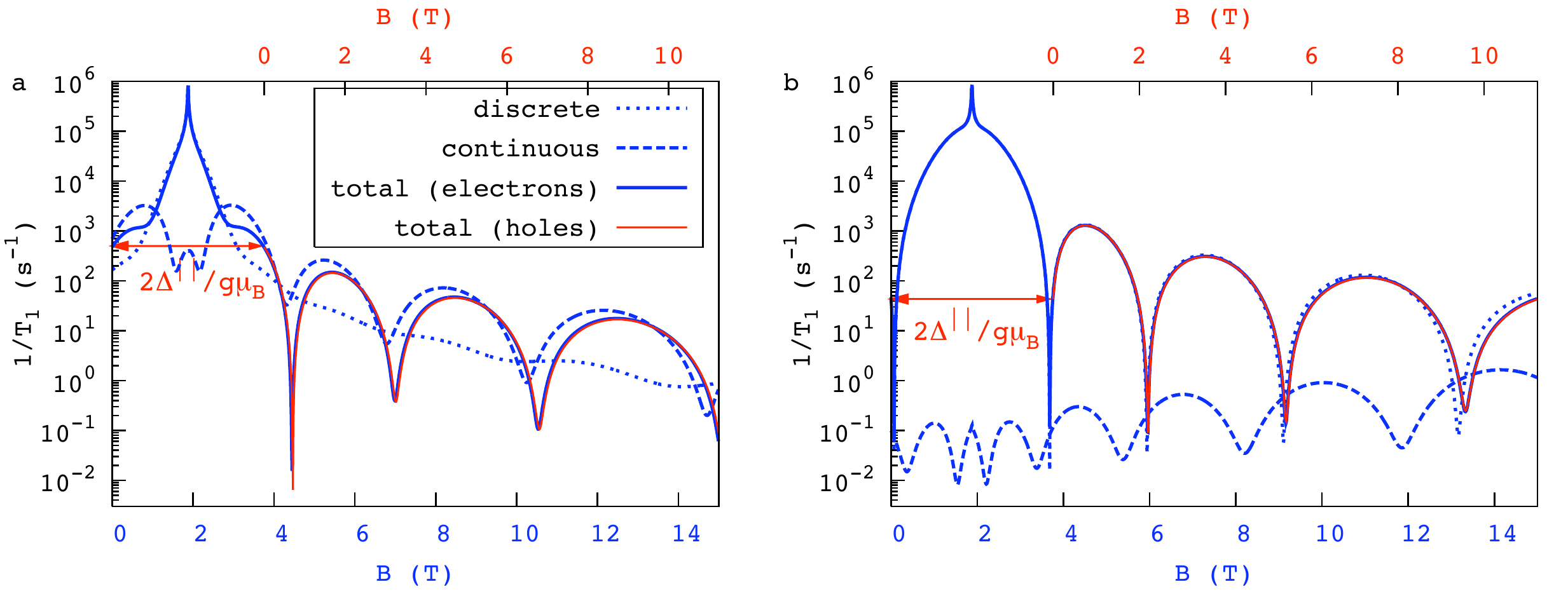}
\end{center} 
	\caption{ Spin relaxation of a (40,0) zigzag NT. The
dependence of  the electron and hole spin relaxation rates $1/T_1$ on a parallel magnetic field $B$ (lower horizontal axis for electrons and upper one for holes)
due to SOI of the two lowest states
($\Psi_{\varkappa_0^\pm,k_0,\pm1/2}$) to higher states of the discrete spectrum
($\Psi_{\varkappa_{\mp1}^\mp,k_n,\mp1/2}$) is shown by the dashed curve and
that due to SOI to states of the continuous spectrum
($\Psi_{\varkappa_{\mp1}^\mp,k,\mp1/2}$) by the dotted curve. The total spin
relaxation rate (solid curves) exhibits destructive interference at each odd
crossing of the curves related to the abovementioned contributions
($R\approx1.6\:$nm, $L=100\:$nm, $g=2$, $T=0.1\:$K,  $V_g=\pm\hbar v
/150R\approx\pm2.3\:$meV {\bf (a)}, and $V_g=\pm\hbar v /3R\approx\pm113\:$meV {\bf
(b)}, where the upper/lower sign is for electrons/holes). The electron (blue
solid curve) and the hole (red solid curve) spin relaxation time coincide by a
shift along the magnetic field axis by $\Delta
B=2\Delta_\mathrm{curv}^\parallel/g\mu_\mathrm{B}$. In Fig.~\ref{fig:One}a (small $V_g$), the continuous spectrrum substantially influences $1/T_1$. In contrast, in Fig.~\ref{fig:One}b (large $V_g$ with many bound states in the dot), the discrete spectrum mainly determines the magnetic field dependence of $1/T_1$.}   \label{fig:One} \end{figure*} 

As shown in Fig.~\ref{fig:One}, the magnetic-field dependence of the spin
relaxation rate of a NT QD is exceptional in comparison to that of a
conventional semiconducting QD. First, there is a singularity
of the electron spin relaxation rate  at $\omega_0\to0$ (or at
$\omega_\mathrm{Z}\to2\tau_3\Delta_\mathrm{curv}^\parallel$)  in contrast to the usual super-Ohmic behavior of $1/T_1$ in
GaAs or InAs QDs (compare Fig.~\ref{fig:One} with Fig.~1 in \onlinecite{GKL}). Remarkably, the position of this symmetric singularity gives us a direct measurement
of the SOI constant $\Delta_\mathrm{curv}^\parallel$ and valley index
$\tau_3$ of an electron in a NT. The singularity is at positive magnetic
fields for the $\mathbf{K}$ point ($\tau_3=1$) and at negative magnetic 
fields for the $\mathbf{K}'$ point ($\tau_3=-1$). In Fig.~\ref{fig:One}, 
where $\tau_3=1$ and
$2\Delta_\mathrm{curv}^\parallel\approx0.22\:$meV, the singularity is at
$B\approx1.9\:$T.  If the SOI constants and $g$ factors are the same for both
electrons and holes, then the electron and hole spin relaxation curves map
onto each other by a shift along the magnetic field axis by $\Delta
B=2\Delta_\mathrm{curv}^\parallel/g\mu_\mathrm{B}$ (compare the blue and the red curves in
Fig.~\ref{fig:One}). 

We have also studied the chirality dependence of the spin relaxation rate as a
function of the magnetic field. Different chirality nanotubes show qualitatively similar spin
relaxation properties.
In Fig.~\ref{fig:Two}, the spin relaxation
 time for NT QDs with different chirality but approximately the same NT radius
 is shown. From this figure we conclude that $T_1$ depends on the chirality of
 a NT, although it has the same qualitative behavior as  a function of a
 magnetic field.

\section{Interference effects in spin relaxation}
\label{Interference}

We note here that the spin relaxation rate for flat GaAs QDs 
in in-plane magnetic
fields is a monotonic function of $B$ (up to about 14T),\cite{GKL} whereas, as shown in
Fig.~\ref{fig:One}, it oscillates with $B$ for NT QDs.
 The oscillations are caused by interference effects of
two types: (i) interference of a {\it phonon wave} in a NT
{\it electron} cavity bounded by the confining potential $V(\zeta)$
due to top gates (see Fig.~\ref{fig:NT}).; (ii) interference between various contributions to the
spin-flip transitions. For clarity, we will now study these two types of
interference phenomena separately.

\subsection{Interference of phonon waves}

\begin{figure}[htb]
\begin{center}
\includegraphics[width=10cm]{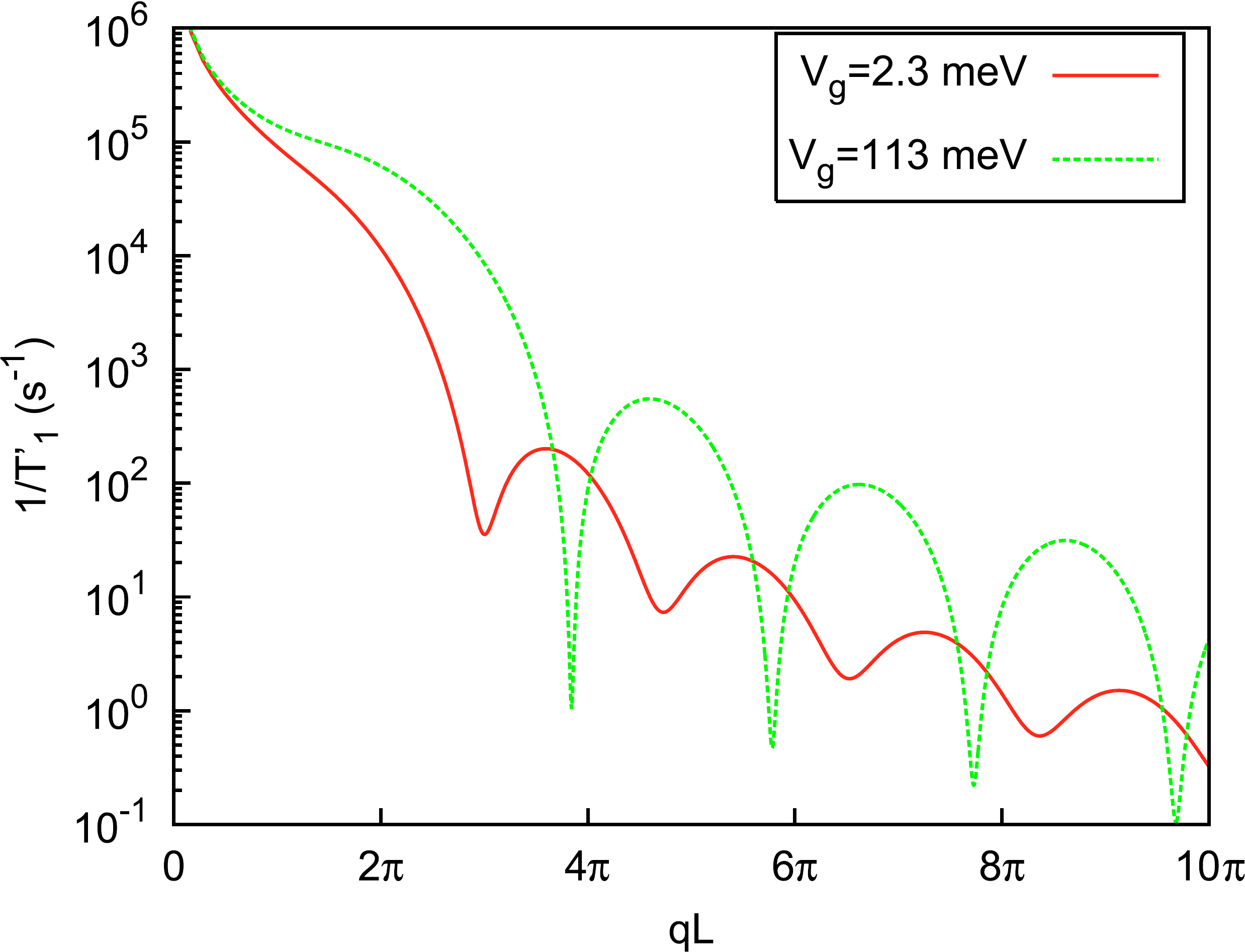}
\end{center}
\caption{Interference phenomena in the spin relaxation rate due to the first contribution to the spin-flip transition ($\propto M_{\omega_\mathrm{0}}'$). Here, $q$ is the phonon wave vector and $L$ is the length of the NT QD ($R\approx1.6\:$nm and $T=0.1\:$K).}
\label{fig:FP}
\end{figure}

To illustrate the first effect, we only consider one contibution to spin-flip
transitions,  namely, that due to the first term in Eq.~(\ref{eq:M_wB}):
\begin{equation}
	M_{\omega_\mathrm{0}}'=\lambda_{k_0}^-\langle \Psi_{\varkappa_0^-,k_0,-1/2}|V_\mathrm{el-ph}|\Psi_{\varkappa_{-1}^-,k_0,-1/2}\rangle.
	\label{eq:Oneterm}
\end{equation} Note that $\lambda_{k_0}^+=0$ due to selection rules (see Eq.~(\ref{eq:selrules})).
The corresponding spin relaxation rate $1/T_1'$ due to this term only is shown
in Fig.~\ref{fig:FP}. 

From Fig.~\ref{fig:FP}, we see that $1/T_1'$ exibits oscillations as a function
of the ratio between the NT QD length $L$ and the phonon wavelength
$\lambda_\mathrm{ph}$: $qL=2\pi (L/\lambda_\mathrm{ph})$. We attribute such
oscillations to interferences of the phonon wave in a NT
electron cavity bounded by the confining potential $V(\zeta)$ due to the top
gates (see Eq.~(\ref{eq:Vg}) and Fig.~\ref{fig:NT}). Such an interference
effect is reminiscent of a Fabry -- Perot -- type 
interference of a phonon wave where the electron levels in the dot play the
role of a cavity.  The coupling between
the phonon wave and the cavity is described by the electron-phonon interaction
$V_\mathrm{el-ph}$.  

%In a conventional optical Fabry -- Perot etalon, light transmission through a
%transparent plate with two reflecting surfaces oscillates as a function of the
%phase difference $\delta$ between two transmitted beams (one of the beams is
%transmitted without reflection and the another one is reflected twice). Note
%that $\delta\propto L/\lambda$, where $L$ is the thickness of the etalon and
%$\lambda$ is the light wavelength.  
%Maximum transmission occurs when the optical path length difference between
%each transmitted beam is an integer multiple of the wavelength.  

%In our case, acoustic vibrations play the role of the light beam and the Fabry
%-- Perot etalon is given by the NT electron cavity bounded by top
%gates. Fig.~\ref{fig:FP} shows the resulting oscillatory behaviour of the
%phonon-induced spin-flip transitions. 
%Minima of such transitions (the spin-flip transition is related to absorption
%of the phonon wave) correspond to maxima of the transmission of the light
%through the optical Fabry -- Perot etalon. 

At the minima in Fig.~\ref{fig:FP}, the coupling
between the electron cavity and the phonon waves becomes small. For an ideal
cavity (with no loss), the matrix element of the spin-flip transition
goes to zero at the minima. For instance, in the case of a rectangular 
hard wall potential, the squared modulus of the phonon-induced spin-flip
transition is given by  
\begin{equation}
	\left|\left\langle\sqrt{\frac2L}\sin\frac{\pi x}{L}\right|e^{iqx}\left|\sqrt{\frac2L}\sin\frac{\pi x}{L} \right\rangle\right|^2=\left(\frac{8\pi^2\sin(qL/2)}{qL[4\pi^2-(qL)^2]}\right)^2,
	\label{eq:HardWall}
\end{equation} 
which is zero at $qL=4\pi,\ 6\pi,\ 8\pi,\ldots$. Therefore, electron-phonon
coupling is switched-off at these interference minima. In the
case of a NT QD with a rectangular confining potential with finite barriers, however, 
due to the penetration of the electron wave function into
classically forbidden region, the electron-phonon coupling is small but nonzero
at the minima of the matrix element of the phonon-induced transition and the
minima are shifted from those for an ideal cavity. As can be
seen from Fig.~\ref{fig:FP}, this shift and the minimal values of the
electron-phonon coupling are more pronounced with increasing the barriers hight $V_g$.  

Interference effects in a NT QD occur only for confinement with well-defined length (for all bound states) and are absent for soft potentials such as parabolic confinement.
Note however that the rectangular potential seems to be a good approximation for
the confinement in a gated NT QD, since Fabry -- Perot interferences (for
electrons) have been observed in such a system.\cite{Liang2001}

\subsection{Coherence of different contributions to spin-flip process}

\begin{figure}[htb]
\begin{center}
\includegraphics[width=10cm]{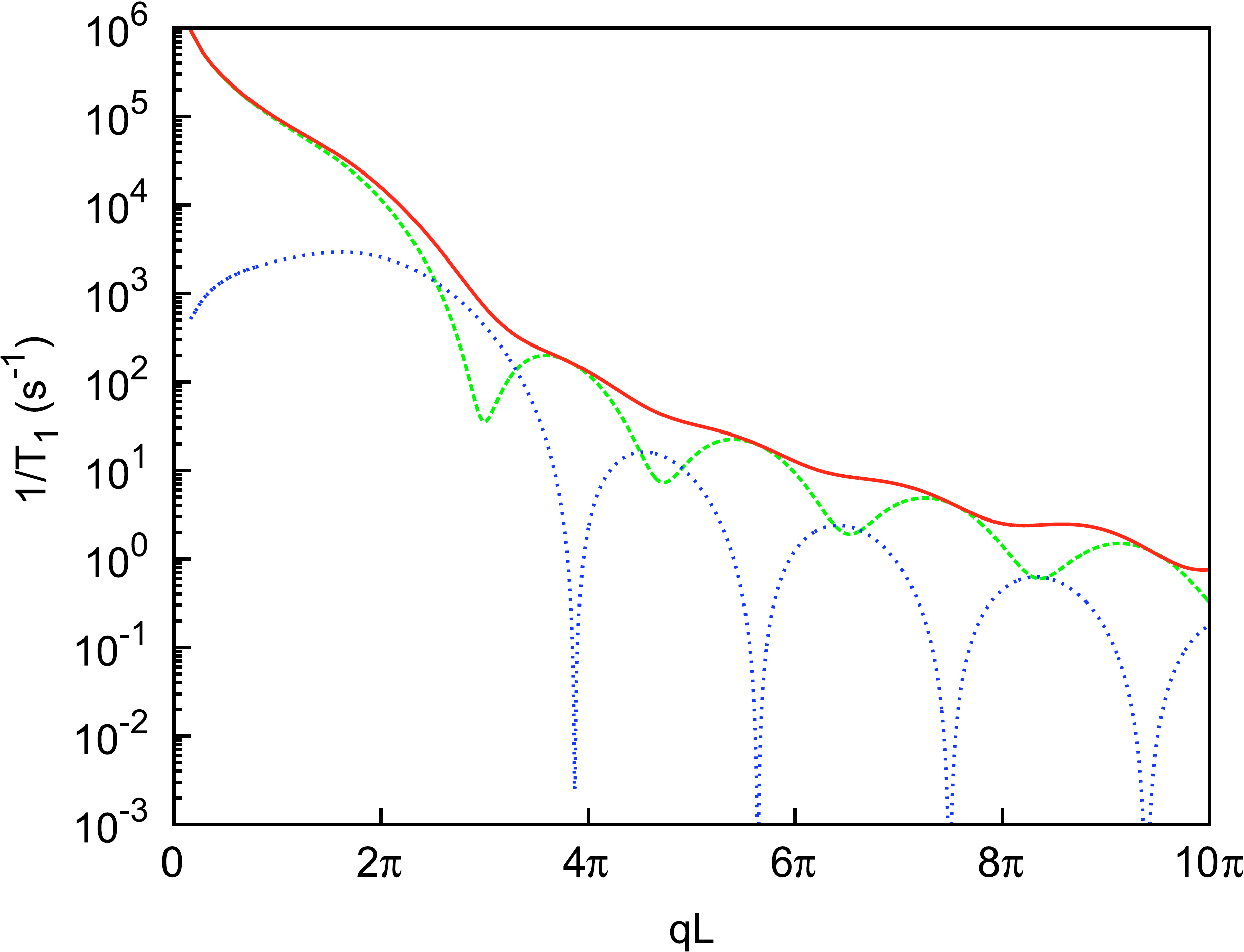}
\end{center}
\caption{Spin relaxation rate due to $M_d^+$ (dashed curve) and $M_d^-$ contribution (dotted curve) to spin-flip transition (see Eq.~(\ref{eq:M_wB2})).
The sum of these two contributions is plotted by solid curve ($R\approx1.6\:$nm, $g=2$, $T=0.1\:$K,  $V_g=\hbar v
/150R\approx2.3\:$meV).
}
\label{fig:Contrib}
\end{figure}

In this subsection, we study interference effects due to various contributions
to the spin-flip transitions described by Eq.~(\ref{eq:M_wB}). Let us consider
the case of weak confinement with small $V_g=2.3\:$meV. In this case,
Eq.~(\ref{eq:M_wB}) can be rewritten as follows: 
\begin{eqnarray}
\label{eq:M_wB2}
M_{\omega_\mathrm{B}}&=&M_d^++M_d^-+M_c^++M_c^-,\\
\nonumber
M_d^+&=& (\lambda_{k}^+)^*\langle \Psi_{\varkappa_1^+,k_0,1/2}|V_\mathrm{el-ph}|\Psi_{\varkappa_0^+,k_0,1/2}\rangle\\
\nonumber
M_d^-&=& \lambda_{k}^-\langle \Psi_{\varkappa_0^-,k_0,-1/2}|V_\mathrm{el-ph}|\Psi_{\varkappa_{-1}^-,k_1,-1/2}\rangle,\\
\nonumber
M_c^+&=&\frac{1}{\pi}\int_{k_c^+}^{\infty} dk (\lambda_{k}^+)^*\langle \Psi_{\varkappa_1^+,k,1/2}|V_\mathrm{el-ph}|\Psi_{\varkappa_0^+,k_0,1/2}\rangle,\\
\nonumber
M_c^-&=&\frac{1}{\pi}\int_{k_c^-}^{\infty} dk \lambda_{k}^-\langle \Psi_{\varkappa_0^-,k_0,-1/2}|V_\mathrm{el-ph}|\Psi_{\varkappa_{-1}^-,k,-1/2}\rangle.
\end{eqnarray}
($k_c^\pm=(|V_g|/\hbar v)\sqrt{1+2\hbar v|\varkappa_{\pm 1}^\pm|/V_g}$).
Here $M_d^+$ and $M_d^-$ are contributions to the spin-flip transitions due to
SOI of the two lowest levels ($\mathcal{E}_{0,0,\pm1/2}$) and
higher \textit{discrete} levels ($\mathcal{E}_{1,0,+1/2}$ and
$\mathcal{E}_{-1,1,-1/2}$). Note that the coupling to other higher levels is
forbidden by the selection rule Eq.~(\ref{eq:selrules}). The
contribution of these two terms to the spin relaxation rate is shown in
Fig.~\ref{fig:Contrib}. It can be seen that these two terms interfere
(constructively) which leads to a change in the amplitude and period of the
oscillations. Such constructive interference (see Fig.\ref{fig:Contrib}) between  $M_d^+$ and $M_d^-$ just increases the spin relaxation rate and, therefore, is not the dominant one. But next we consider a different interference effect which reduces $1/T_1$ by several orders of magnitude.

\subsubsection{Destructive interference}

The remaining and most intriguing interference effect is the one between $M_d^+$ and $M_c^+$ (or between $M_d^-$ and $M_c^-$). These terms are a generated by SO coupling  the two lowest states to the excited discrete and the continuous spectrum, respectively. From Fig.~\ref{fig:One}a and Fig.~\ref{fig:Three} we see that these contributions at some magnetic field interfere destructively leading to a strong increase of the spin relaxation time up to 4 orders of magnitude. Strikingly, such destructive interference is robust against a change of parameters, although being most evident when the terms $M_d^\pm$ and $M_c^\pm$ have comparable contributions to the spin-flip transitions (compare Figs.~\ref{fig:One}a and \ref{fig:One}b).

Let us give a physical explanation for this phenomenon. First of all we note that the diagonal elements of the electron-phonon coupling ($\propto g_1$) (see Eq.~(\ref{eq:B}) for details) give the main contribution to the spin-flip transitions with respect to the non-diagonal ones ($\propto g_2$), since $g_1\gg g_2$. As a result, the destructive interference occurs due to diagonal elements of $V_\mathrm{el-ph}$ and the elements $\propto g_2$ just modulate the strength of the effect, i.e., the depth of the dips in the spin relaxation curve. Therefore, in this subsection, we consider diagonal electron-phonon coupling ($\propto g_1$) only.

The terms of the spin-flip matrix element due to coupling to the first exited subband with $m=1$ (see Eq.~(
\ref{eq:M_wB2})) can be written as follows:
\begin{eqnarray}
	M_d^+&=&  (\lambda_{k}^+)^*\langle \Psi_{\varkappa_1^+,k_0,1/2}|V_\mathrm{el-ph}|\Psi_{\varkappa_0^+,k_0,1/2}\rangle\propto
	\int_{-\infty}^\infty d\zeta e^{iq_0\zeta}\Phi_{\varkappa_1^+,k_0}^\dagger(\zeta)\Phi_{\varkappa_0^+,k_0}(\zeta),\\
	\label{eq:Mcp}
M_c^+&=&\frac{1}{\pi}\int_{k_c^+}^{\infty} dk (\lambda_{k}^+)^*\langle \Psi_{\varkappa_1^+,k,1/2}|V_\mathrm{el-ph}|\Psi_{\varkappa_0^+,k_0,1/2}\rangle\propto	\int_{-\infty}^\infty d\zeta e^{iq_0\zeta}\int_{k_c^+}^{\infty} dk (\lambda_{k}^+)^*\Phi_{\varkappa_1^+,k}^\dagger(\zeta)\Phi_{\varkappa_0^+,k_0}(\zeta)
\end{eqnarray}
Note that $\Phi_{\kappa_0^+,k_0}^\dagger(\zeta)\Phi_{\kappa_0^+,k_0}(\zeta)$ is a symmetric function of $\zeta$ with respect to the center of the NT QD ($\zeta=L/2$). In the dot area ($0\le\zeta\le L$), it might be approximated by a function $\propto\cos(\tilde{k}_0 (L/2-\zeta) )$ with exponential tails in the classically forbidden areas ($\zeta<0$ and $\zeta>L$). In addition to the selection rule Eq.~(\ref{eq:selrules}), it is easy to find that $\langle\Phi_{\kappa_{m'}^\pm,k_{n'}}(\zeta)|\Phi_{\kappa_m^\pm,k_n}(\zeta)\rangle=1+\mathrm{sgn}(m'-1/3)\mathrm{sgn}(m-1/3)(-1)^{n'+n}$, because $\Phi_{\kappa_{m'}^\pm,k_{n'}}^\dagger(\zeta)\Phi_{\kappa_m^\pm,k_n}(\zeta)$ is either odd or even with respect to inversion at $\zeta=L/2$. Therefore, $M_d^+=0$ at $q_0=0$, since $\Phi_{\varkappa_1^+,k_0}^\dagger(\zeta)\Phi_{\varkappa_0^+,k_0}(\zeta)$ is an asymmetric function with respect to $\zeta=L/2$ at which it has a node. Thus, $\Phi_{\varkappa_1^+,k_0}^\dagger(\zeta)\Phi_{\varkappa_0^+,k_0}(\zeta)$ is found to be well approximated by a function $\sin(k'(L/2-\zeta))$ defined at $0\le\zeta\le L$. Now we consider Eq.~(\ref{eq:Mcp}). After integration over $k$, we could assume that the dependence of $\Phi_{\varkappa_1^+,k}$ on $\zeta$ is integrated out, therefore, $M_c^+$ is a symmetric function of $\zeta$ with respect to $\zeta=L/2$, which we approximate by $-i\cos(k''(L/2-\zeta))$ defined at $0\le\zeta\le L$. Using these assumptions, we get the following estimations:
\begin{eqnarray}\label{eq:f0}
M_d^+& \propto & \int_0^L d\zeta e^{iq_0\zeta}\sin(k'(L/2-\zeta))=f_1(q_0L)+if_2(q_0L),\\
M_c^+& \propto & \int_0^L d\zeta e^{iq_0\zeta}i\cos(k''(L/2-\zeta))=f_3(q_0L)+if_4(q_0L),\\
\label{eq:f1}
f_1(q_0L)&=& \int_0^L d\zeta \cos{q_0\zeta}\sin(k'(L/2-\zeta))=2\frac{-k'\cos\frac{k'L}{2}\sin\frac{q_0L}{2}+q_0\sin\frac{k'L}{2}\cos\frac{q_0L}{2}}{(k')^2-q_0^2}\sin\frac{q_0L}{2},\\
\label{eq:f2}
f_2(q_0L)&=& \int_0^L d\zeta \sin{q_0\zeta}\sin(k'(L/2-\zeta))=2\frac{k'\cos\frac{k'L}{2}\sin\frac{q_0L}{2}-q_0\sin\frac{k'L}{2}\cos\frac{q_0L}{2}}{(k')^2-q_0^2}\cos\frac{q_0L}{2},\\
\label{eq:f3}
f_3(q_0L)&=& \int_0^L d\zeta \sin{q_0\zeta}\cos(k''(L/2-\zeta))=2\frac{k''\sin\frac{k''L}{2}\cos\frac{q_0L}{2}-q_0\cos\frac{k''L}{2}\sin\frac{q_0L}{2}}{(k'')^2-q_0^2}\sin\frac{q_0L}{2},\\
\label{eq:f4}
f_4(q_0L)&=& -\int_0^L d\zeta \cos{q_0\zeta}\cos(k''(L/2-\zeta))=2\frac{-k''\sin\frac{k''L}{2}\cos\frac{q_0L}{2}+q_0\cos\frac{k''L}{2}\sin\frac{q_0L}{2}}{(k'')^2-q_0^2}\cos\frac{q_0L}{2}.
\end{eqnarray}

\begin{figure}[htb]
\begin{center}
		\includegraphics[width=18cm]{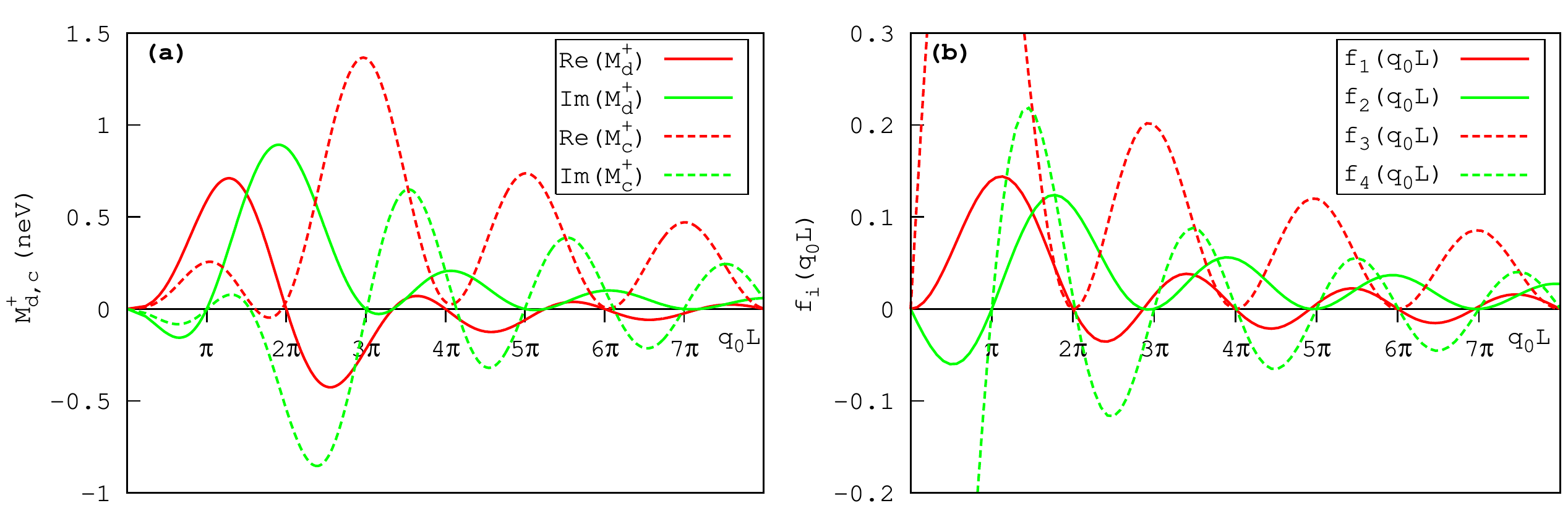}
\end{center}
\caption{(a) Dependence of the real and imaginary parts of matrix elements $M_d^+$ and $M_c^+$ (see Eq.~(\ref{eq:M_wB2}) for details) on the ratio of the NT length and phonon wavelength $q_0L$ ($R\approx1.6\:$nm, $l=100\:$nm, $g=2$). (b) Approximation of the previous dependence by $f_i(q_0L)\ (i=1,...,4)$ ($k'L=k''L=0.7$) (see Eqs.~(\ref{eq:f0})--(\ref{eq:f4})).
}
\label{fig:Interf}
\end{figure}
We have plotted the functions $f_i(q_0L)$ (see Fig.~\ref{fig:Interf}b) in comparison to the real and imaginary parts of $M_{c,d}^+$ (Fig.~\ref{fig:Interf}a). There is a good agreement between the corresponding functions (plotted with the same line style) except for the region of $q_0L<2\pi$ for $f_3(q_0L)$ and $f_4(q_0L)$. From Fig.~\ref{fig:Interf}, Eqs.~(\ref{eq:f1}) and (\ref{eq:f2}), one can see that $\mathrm{Re}(M_d^+)$ and $\mathrm{Im}(M_d^+)$ have zeroes at $k'\tan q_0L/2=q_0\tan k'L/2$ (which are close to  $q_0L=(2n+1)\pi$, where $n=1,2,3,\ldots$), in addition, $\mathrm{Re}(M_d^+)$ is zero at $q_0L=2\pi n$ and $\mathrm{Im}(M_d^+)$ is zero at $q_0L=(2n-1)\pi$. From Fig.~\ref{fig:Interf}, Eqs.~(\ref{eq:f3}) and (\ref{eq:f4}), we get that $\mathrm{Re}(M_c^+)$ and $\mathrm{Im}(M_c^+)$ have zeroes at $k''\tan k''L/2=q_0\tan q_0L/2$ (which are close to  $q_0L=2n\pi$), in addition, $\mathrm{Re}(M_c^+)$ is zero at $q_0L=2\pi n$ and $\mathrm{Im}(M_c^+)$ is zero at $q_0L=(2n-1)\pi$. Zeroes of the above functions determine the regions of $q_0L$ in which the sign of those functions is constant, namely,
\begin{eqnarray}
	\mathrm{Re}(M_d^+),\ f_1(q_0L)\le0&&\ \mathrm{for}\ 2n\pi\le q_0L\lesssim (2n+1)\pi,\\
	\mathrm{Re}(M_c^+),\ f_3(q_0L)>0&&\ \mathrm{almost\ for\ any}\ q_0L,\\
	\mathrm{Im}(M_d^+),\ f_2(q_0L)>0&&\ \mathrm{almost\ for\ any}\ q_0L>\pi,\\
	\mathrm{Im}(M_c^+),\ f_4(q_0L)\le0&&\ \mathrm{for}\ 2n\pi\lesssim q_0L\le (2n+1)\pi.
	\label{eq:regions}
\end{eqnarray}
From these equations we find that for $2n\pi\lesssim q_0L\lesssim(2n+1)\pi$ the functions $\mathrm{Re}(M_d^+)$ and  $\mathrm{Re}(M_c^+)$, as well as, $\mathrm{Im}(M_d^+)$ and  $\mathrm{Im}(M_c^+)$ have opposite signs. In other words, \textit{due to odd or even symmetry of vector states with respect to  the center of a NT QD ($\zeta=L/2$), the terms of spin-flip transitions $M_d^+$ and $M_c^+$ combine in antiphase at $2n\pi\lesssim q_0L\lesssim(2n+1)\pi$, resulting in destructive interference of those contributions to the spin relaxation rate.} Note that $M_{d,c}^-$ terms have similar behaviour and the same statements hold true for those. 
\begin{figure}[htb]
\begin{center}
\includegraphics[width=10cm]{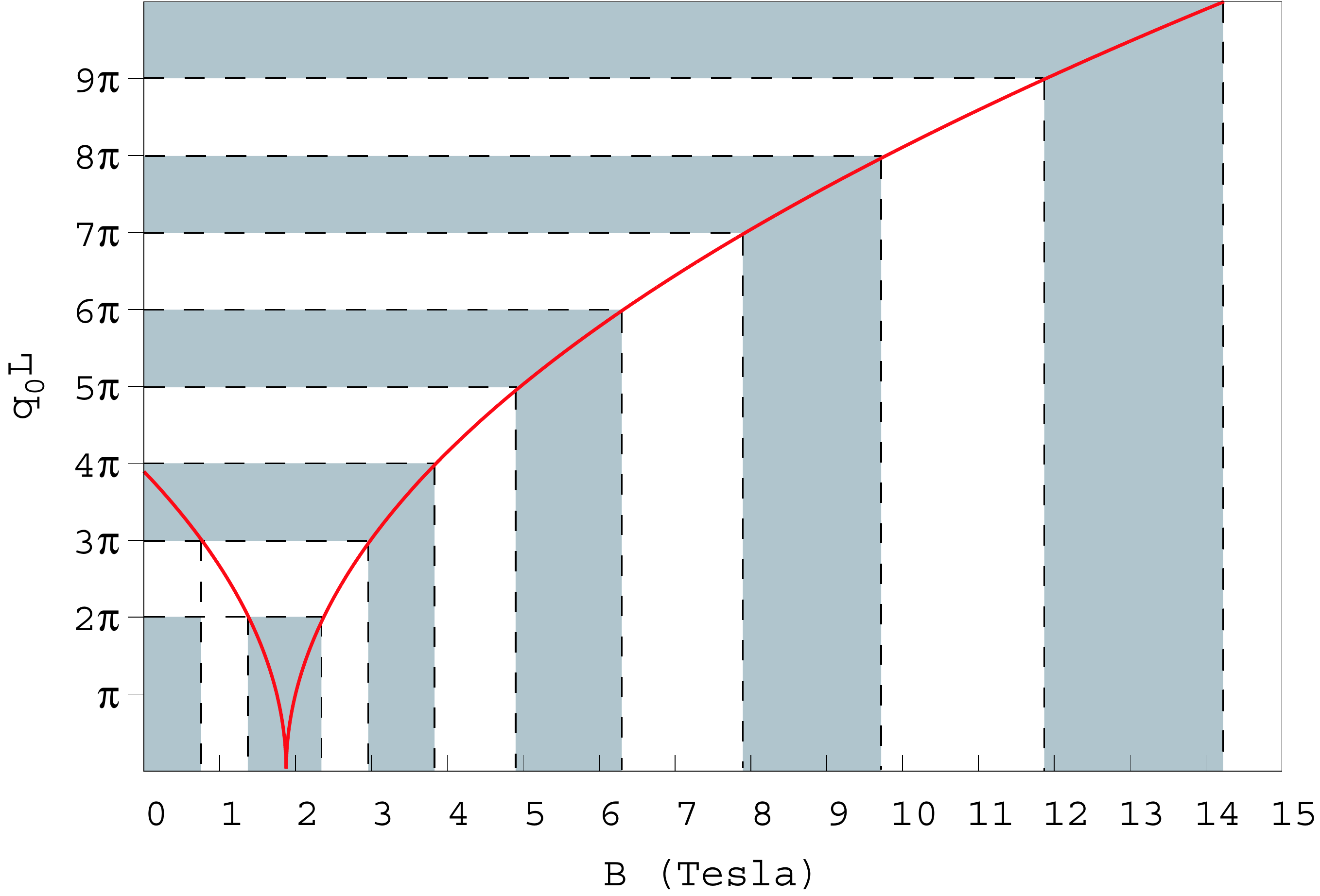}
\end{center}
\caption{The dependence of $q_0L$ on  magnetic field $B$ ($R\approx1.6\:$nm, $g=2$, $\Delta^\parallel_\mathrm{curv}=0.1085\:$meV). Non-shaded areas stand for regions of magnetic field where destructive interference occurs.
}
\label{fig:Area}
\end{figure}
Here $q_0=\sqrt{\sqrt{2}|\hbar\omega_\mathrm{Z}-2\tau_3\Delta_\mathrm{curv}^\parallel|/\hbar c_\mathrm{S}R}$ is the phonon wave vector of the resonant spin-flip transition. The magnetic-field dependence of $q_0$ is shown in Fig.~\ref{fig:Area} (red curve). The regions of $q_0L$ and $B$, where the destructive interference is expected, are shown by non-shaded areas. From Figs.~\ref{fig:One}a, \ref{fig:Two}, and \ref{fig:Three}, we see that the destructive interference dips in the spin relaxation rate are in the defined regions shown in   Fig.~\ref{fig:Area}.

 The effect is stronger for smaller $V_g$, when the number of discrete levels
 is lower (see Fig.~\ref{fig:One}a). In this case, the spin relaxation is
 predominantly due to coupling to the continuous spectrum. With increasing the
 voltage $V_g$ applied to top gates, the number of discrete levels and the
 spacing between the ground state and the lower bound of the continuous
 spectrum increases. (For instance, in the
 case of $V_g=2.3\:$meV, there are only two discrete levels, while, for
 $V_g=113\:$meV, there are about 15 quantized levels in each subband.)
This decreases (increases) the contribution of the
 continuous (discrete) spectrum to the spin relaxation rate  and increases the
 total spin relaxation rate (compare Fig.~\ref{fig:Three}, where $V_g\approx8.5\:$meV, with Fig.~\ref{fig:One}a, where $V_g\approx2.3\:$meV, for the same type of NT QD). Such rich and unexpected behavior of
the  spin relaxation in NT QDs is remarkable and opens up broad perspectives for spintronics in
 carbon nanostructures.

%The second distinction is that the overall behavior of $1/T_1$ tends to
%decrease with $B$ for NT QDs, while it increases with $B$ for conventional
%QDs. 

\section{Conclusions}
  In conclusion, contrary to the common believe
 that spin-orbit interaction is weak and insignificant in carbon
materials, we have shown that the situation is actually much richer and that spin-orbit
interaction can be very important in nanotubes.
We have studied spin relaxation and decoherence caused by electron-lattice and
spin-orbit interaction and predict striking non-monotonic effects induced by magnetic fields
$B$. For particular values
of $B$, destructive interference occurs resulting in ultralong spin
relaxation  times $T_1$ exceeding tens of seconds. For small phonon frequencies $\omega$, we find a
$1/\sqrt{\omega}$ spin-phonon noise spectrum --
a novel dissipation channel for spins in quantum dots -- which can reduce
$T_1$ by many orders of magnitude. We show that
nanotubes exhibit zero-field level splitting caused by spin-orbit
interaction. This enables an all-electrical and phase-coherent control of
spin -- the hallmark of spintronics.

We thank G.~Burkard, S.~Ilani, L.~Kouwenhoven, and 
L.~Vandersypen for useful discussions and 
 E.~Klinovaya for pointing out several typos.
 We acknowledge
support from the Swiss NSF, NCCR Nanoscience, ONR, and JST ICORP.
\appendix 
%--+--

\section{Spin splitting and electric-dipole spin resonance at zero magnetic fields} 
\label{EDSR}
\begin{figure}[thb]
\begin{center}
\includegraphics[width=10cm]{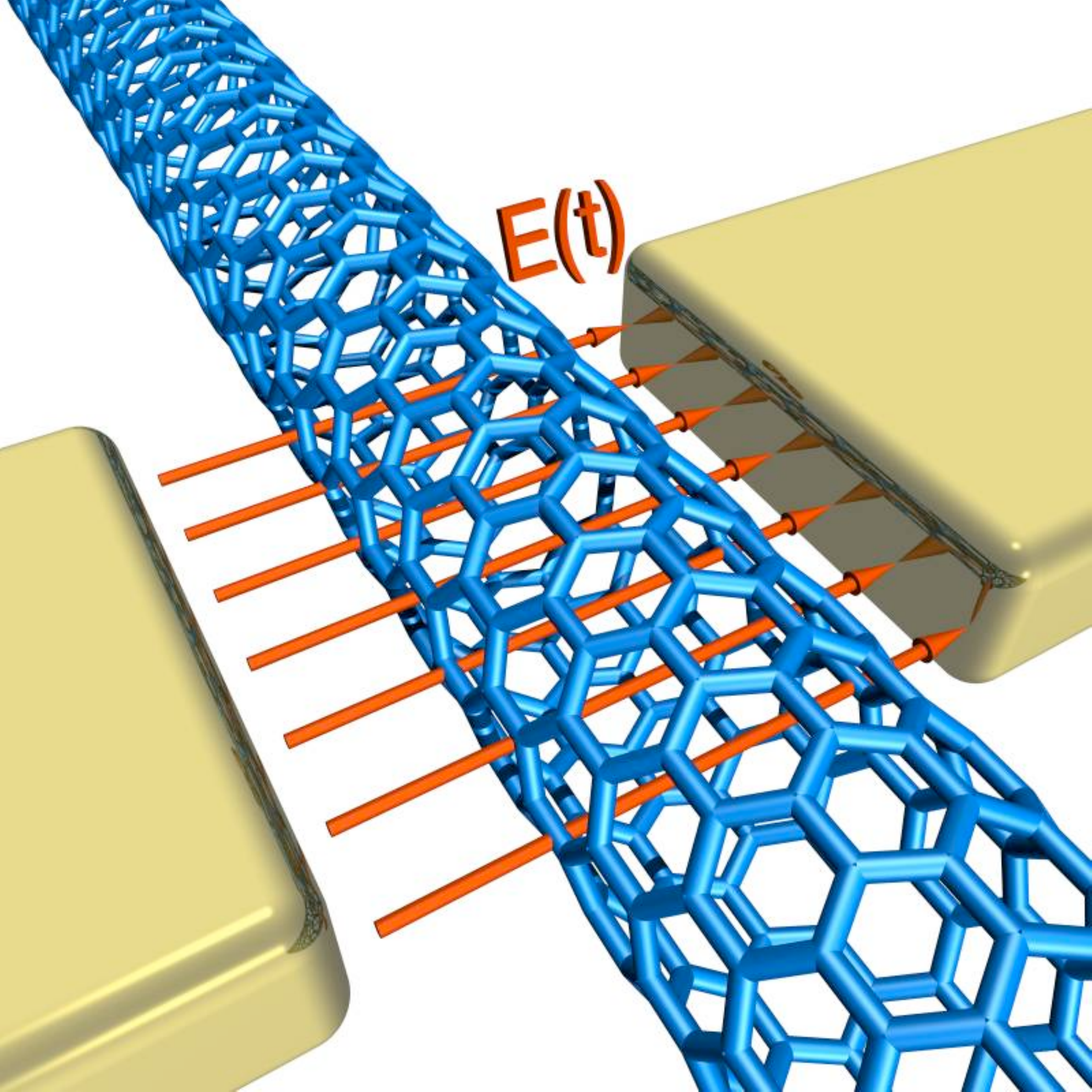}
\end{center}
\caption{Scheme of a NT with two electric side gates. An applied ac voltage between the gates creates an oscillating electric field perpendicular to the NT axis. Such a setup enables electrically-induced coherent spin manipulation, due to zero-field spin splitting in semiconducting NTs.}
\label{fig:El_gates}
\end{figure}

From Eq.~(\ref{eq:energy}), one can find that there is a zero-field splitting between spin-up and spin-down states:
\begin{eqnarray}\nonumber
\left.\left(\mathcal{E}_{0,0,1/2}-\mathcal{E}_{0,0,-1/2}\right)\right|_{B=0}&=& \hbar v \sqrt{(-\tau_3 \nu/3R+\Delta_\mathrm{curv}^\parallel/\hbar v)^2+k_0^2}
-\hbar v \sqrt{(-\tau_3\nu/3R-\Delta_\mathrm{curv}^\parallel/\hbar v)^2+k_0^2}\\
&\approx&-2\tau_3\nu\Delta_\mathrm{curv}^\parallel,
\label{eq:ZFS}
\end{eqnarray}
where $\nu=0,\pm1$, and where we have taken into account that $k_0\ll1/R$ and
neglected intervalley mixing. Due to the first term in
Eq.~(\ref{eq:HSO_curv}), there is spin mixing and, therefore, coupling between
corresponding states (see also Eq.~(\ref{eq:psi})) which allows electric 
dipole transitions between them. Consider an oscillating electric field (see Fig.~\ref{fig:El_gates}):
$\mathbf{E}(t)=E\mathbf{e}_\perp\sin\omega t$, $\mathbf{e}_\perp$ is the unit vector perpendicular to the NT axis. An interaction between the
electric field and an electron in a NT, which leads to electric-dipole
transitions, is given by the following operator: 
\begin{equation}
H_E=\frac{|e|E}{m_0\omega}\cos\omega tP_\perp=\frac{-i|e|\hbar E}{m_0R\omega}\cos\omega t\sin\varphi\frac{\partial}{\partial\varphi},
\label{eq:H_E}
\end{equation}
where  $m_0$ is the bare electron mass and $P_\perp=-i\hbar\sin\varphi\partial_\varphi/R$ is the electron momentum along $\mathbf{e}_\perp$. Here we assume that the influence of the lattice potential can be neglected for the estimation of the electric-dipole transitions.
Therefore, using Eq.~(\ref{eq:psi}), the matrix element of the electric-dipole transitions can be expressed as
\begin{eqnarray}\nonumber
\langle\psi_{0,0,+1/2}|H_E|\psi_{0,0,-1/2}\rangle&=& \frac{-i|e|\hbar E}{2m_0R\omega}\cos\omega t
\sum_k\left[\left(\lambda_k^-\right)^*\langle\Psi_{\varkappa_{-1}^-,k,-1/2}|ie^{-i\varphi}\frac{\partial}{\partial\varphi}|\Psi_{\varkappa_{0}^-,k_0,-1/2}\rangle\right.\\
&&+\left.\lambda_k^+\langle\Psi_{\varkappa_{0}^+,k_0,+1/2}|ie^{-i\varphi}\frac{\partial}{\partial\varphi}|\Psi_{\varkappa_{+1}^+,k,1/2}\rangle\right]= \hbar\omega_R\cos\omega t,\\
\label{eq:H_EMEl}
\omega_R&=& \frac{i|e|E}{2m_0R\omega}
\sum_k\left[\left(\lambda_k^-\right)^*\langle\Phi_{\varkappa_{-1}^-,k,-1/2}|\Phi_{\varkappa_{0}^-,k_0,-1/2}\rangle+\lambda_k^+\langle\Phi_{\varkappa_{0}^+,k_0,+1/2}|\Phi_{\varkappa_{+1}^+,k,1/2}\rangle\right].
\end{eqnarray}
Here the sum includes summation over the discrete $k_n$ and integration over the continuous $|k|\ge (|V_g|/\hbar v)\sqrt{1+2\hbar v|\kappa_m|/V_g}$. Numerical evaluation leads to the following estimates for
the resonance frequency $\omega=2\Delta^\parallel_\mathrm{curv}/\hbar\approx 33\times10^{10}\:$s$^{-1}$
 ($\Delta^\parallel_\mathrm{curv}\approx0.11\:$meV) and Rabi frequency $\omega_\mathrm{R}\approx 1.6\times10^5\:$s$^{-1}$
 at $E=10\:$V$/\:$cm and $V_g=2.3\:$meV.

 \section{Valley-orbit and spin-orbit interactions}
 \label{valley}
\begin{figure}[thb]
\begin{center}
\includegraphics[width=10cm]{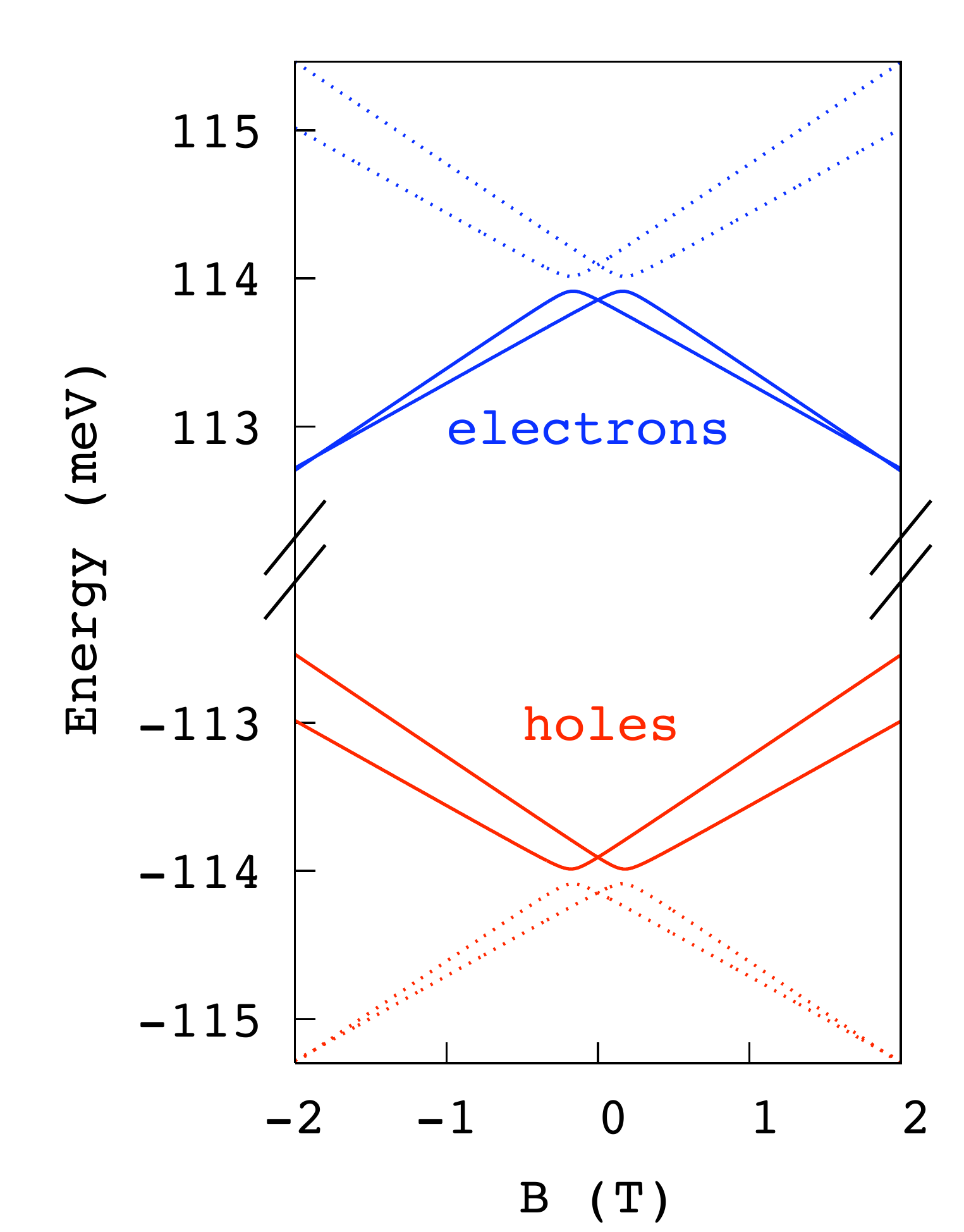}
\end{center}
\caption{  Magnetic field dependence of energy levels for electrons (blue curves) and holes (red curves). Dashed curves are the excited states in a NT QD ($R\approx1.6\:$nm, $L=100\:$nm, $g=2$, $V_g\approx\pm8.5\:$meV (the upper/lower sign is for electrons/holes), $\Delta_\mathrm{curv}^\parallel=0.11\:$meV, $\Delta_\mathbf{K-K'}=0.05\:$meV). }
\label{fig:Spectrum}
\end{figure}

In this section, we consider a particle in a NT described by the
Hamiltonian (\ref{eq:H0}) with the longitudinal confinement (\ref{eq:Vg}) in a
parallel magnetic field. The discrete spectrum of a such system is given by 
\begin{equation}
	E_{m,n,S_\zeta}=\pm \hbar v\sqrt{\kappa_m^2+k_n^2}+S_\zeta\hbar
        \omega_\mathrm{Z}, 
	\label{eq:sptr}
\end{equation}
where $\kappa_m=(m-\tau_3\nu/3+\varphi_\mathrm{AB})/R$ and $\varphi_\mathrm{AB}=\Phi_\mathrm{AB}/\Phi_0$. Each
level is four-fold degenerate (at $B=0$) due to valley and spin
degeneracy. Now, we take SOI into account. For
definiteness, we consider only the second term in Eq.~(\ref{eq:HSO_curv})
which leads to zero-field splitting: 
\begin{equation}
	H^\mathrm{curv}_\mathrm{SO}=\Delta^\parallel_\mathrm{curv}\tau_3\sigma_1 2 S_\zeta.
	\label{eq:HSO2}
\end{equation}
	Moreover, within a minimal model we take intervalley mixing due to non-magnetic impurities or
        structure defects into account. In this case, the intervalley mixing
        can be described by the following term: 
	\begin{equation}
		H_\mathbf{K-K'}=\Delta_\mathbf{K-K'}\tau_1,
		\label{eq:KK'mix}
	\end{equation}
	where $\tau_1$ is the Pauli matrix operating on
			valley-index space.
The eigenvalues of the operator $H_0+V(\zeta)+ H^\mathrm{curv}_{\mathrm{SO}} +
H_\mathbf{K-K'}$ (for $m=0$ subband) are given by 
\begin{eqnarray}\nonumber
	E_{0,n,S_\zeta} &=& \pm\hbar v
        \left[\nu^2/9R^2+k_n^2+k^2_\mathbf{K-K'}+(\varphi_\mathrm{AB}/R+2S_\zeta
          k_\mathrm{SO})^2\right.\\
	  &&\left.+2\beta\sqrt{(\varphi_\mathrm{AB}/R+2S_\zeta
	  k_\mathrm{SO})^2\nu^2/9R^2+k^2_\mathbf{K-K'}(\nu^2/9R^2+k_n^2)}\right]^{1/2}+ S_\zeta\hbar\omega_\mathrm{Z},
	\label{eq:EnFull}
\end{eqnarray}
where $k_\mathbf{K-K'}=\Delta_\mathbf{K-K'}/\hbar v$, $k_\mathrm{SO}=\Delta_\mathrm{curv}^\parallel/\hbar v$.
The  energy spectrum of the lowest electron energy levels and highest hole levels described by Eq.~(\ref{eq:EnFull}) is shown in Fig.~\ref{fig:Spectrum} ($n=0$).
 In Eq.~(\ref{eq:EnFull}), the plus (minus) sign corresponds to electron (hole) energy levels. $\beta=1$ for the upper branch of the energy spectrum (blue dashed and red solid curves) and $\beta=-1$ for the lower branch (blue solid and red dashed curves). 
 Using  $|\kappa_m|\gg k_n$, we rewrite Eq.~(\ref{eq:EnFull}) in the following way:
\begin{eqnarray}
	E_{0,n,S_\zeta} &=& \pm\hbar v\sqrt{\nu^2/9R^2+k_n^2}+\beta\hbar v\sqrt{k^2_\mathbf{K-K'}+(\varphi_\mathrm{AB}/R+2S_\zeta
	k_\mathrm{SO})^2}+ S_\zeta\hbar\omega_\mathrm{Z}+O\left( 3\nu k_nR\varphi_\mathrm{AB} \right).
	\label{eq:EnFull2}
\end{eqnarray}
From this equation we find that the zero-field splitting of levels is given by $\hbar v \sqrt{k^2_\mathrm{SO}+k^2_\mathbf{K-K'}}$ and anticrossings (with the magnitude $2|\Delta_\mathbf{K-K'}|$) occur at $\varphi_\mathrm{AB}=-2S_\zeta k_\mathrm{SO}R=0$. 

Due to intervalley coupling, electron states of different nonequivalent $\mathbf{K}$ points are mixed:
\begin{equation}
	\Psi_\mathbf{K}\approx\Psi_\mathbf{K}^{(0)}+\frac{\Delta_\mathbf{K-K'}}{\sqrt{\Delta_\mathbf{K-K'}^2+(\hbar v \varphi_\mathrm{AB}/R+2S_\sigma\Delta^\parallel_\mathrm{curv})^2}}\Psi_\mathbf{K'}^{(0)}.
	\label{eq:mixing}
\end{equation}
The mixing is maximal at the anticrossing points (at that point, the electron state is just a superposition of those corresponding to different $\mathbf{K}$ points) and suppressed away from them. Therefore,  there is a way to control 
the intervalley mixing for a NT by a magnetic field and it makes NTs attractive for a valley-qubit realization (qubits whose quantum states are defined by the valley index).\cite{TBLB2007,Recher,Carlo}

At anticrossing points, mixing of $\mathbf{K}$ and  $\mathbf{K'}$ valleys is strong and intervalley scattering could occur. Note that such scattering is energetically forbidden for first-order processes (with one-phonon scattering). Indeed, 
scattering from $\mathbf{K}$ to $\mathbf{K'}$ point requires a large change in the electron wave vector ($|\mathbf{K-K'}|=|\mathbf{K}|$), while the energy difference between the scattering states is small ($2\Delta_\mathbf{K-K'}\le0.5\:$meV). Phonons in a NT at the $\mathbf{K}$ point of the phonon dispersion have much higher energy ($\omega_K>600\:$cm$^{-1}$ which correspond to $75\:$meV) \cite{Saito2003} and, therefore, first-order single-phonon intervalley scattering is forbidden. However, Raman spectroscopy has shown  that such an intervalley scattering is allowed for photo-excited electrons. Such transitions are attributed to second-order Raman processes by two phonon emission or emission of one phonon and elastic scattering on lattice defects (so called D- and G'-bands in the Raman spectra).\cite{Saito2003} We assume that similar processes could occur in our case, due to spontaneous 
phonon emission and absorption with $\hbar\omega_1-\hbar\omega_2=2\Delta_\mathbf{K-K'}$ and $\mathbf{q_1-q_2=K}$ or to emission of a single phonon with $\mathbf{q=K}$ and elastic scattering on lattice defects, but are nevertheless less probable.

\end{document}